\documentclass[%
 reprint,
 amsmath,amssymb,
 aps,
prb,
]{revtex4-2}

\usepackage{graphicx}% Include figure files
\usepackage{dcolumn}% Align table columns on decimal point
\usepackage{bm}% bold math
\usepackage{xcolor}

\usepackage{ulem}

\begin{document}

\preprint{APS/123-QED}

\title{Incorporating density jumps and species-conserving dynamics in XPFC binary alloys}

\author{Matthew J. Frick}
\affiliation{Department of Physics, Centre for the Physics of Materials, McGill University, Montreal, QC, Canada}

\author{Nana Ofori-Opoku}
\affiliation{Computational Techniques Branch, Canadian Nuclear Laboratories, Chalk River, ON, Canada}

\author{Nikolas Provatas}
\affiliation{Department of Physics, Centre for the Physics of Materials, McGill University, Montreal, QC,\;Canada}

\begin{abstract}
    This work presents a consistent formulation of the structural phase-field-crystal model of substitutional binary alloys that allows for the description phases of unequal densities, a key feature in solidification. We further develop the dynamics of the model to be consistent with conserved Langevine dynamics in the true governing species densities. Additionally, this work expands on the ability to control pressure, so far only implemented in pure materials, to binary alloys by improving the control system that controls pressure from previous work. We study the equilibrium properties of the new model, and demonstrate that control of pressure can drive various kinematic microscopic processes in materials such as grain boundary pre-melting, phase instability, and grain or inter-phase boundary motion.
\end{abstract}

\maketitle

\section{\label{sec:Intro} Introduction}

The macroscopic material properties of engineering alloys have long been understood to arise out of a complicated relationship to the microstructure that forms during the casting and thermomechanical processing of a material. However, due to the prohibitive difficulty of directly imaging or performing \textit{in situ} measurements during these processes the understanding of the microstructural formation is often empirical.

Advances in the ability to predict microstructure have been largely driven by atomistic and mesoscale phase field type modelling. One such class of models are so-called Phase Field Crystal (PFC) models~\cite{elder2007}. These are phase field models with a periodic order parameter, which allow the resolution of atomic-scale structure and defects that evolve on inherently diffusive time-scales. These class of models thus have many of the complex physical mechanism that need to be built into traditional phase field models such as strain relaxation, elasticity, and arbitrary grain orientation, which arise holistically from the form of a PFC free energy~\cite{grant2004}. The ability to model robust types of crystal structures beyond that possible in the original PFC model has since led to a class of PFC models referred to as Structural Phase Field Crystal (XPFC) models~\cite{greenwood2010, greenwood2011,greenwood2011binary}, the alloy version of which was then expanded upon to model arbitrary enthalpy of mixing by Smith \textit{et al.}~\cite{smith2017}.

A key feature absent from past PFC models has been a consistent description and control of bulk density or volume changes between phases. This is a crucial prerequisite required of any model describing shrinkage and void formation during rapid solidification, as well as free surface creation under stress. Kocher \textit{et al.}~\cite{kocher2015} have recently addressed this issue in pure materials by considering coexistence in pure materials in density($\rho$)-pressure($p$)-temperature($T$) space, and by introducing a gaseous phase to allow a more complete description of possibilities enabled by large bulk volume/pressure changes. The first step toward controlling the  $\rho$-$p$-$T$ space, a vapour forming binary alloy has been attempted by Wang \textit{et al.}~\cite{Wang2017} using a phenomenology that interpolates the free energy of a simple triangular-forming crystal phase with a liquid-vapour system through changes in local bulk density. The dynamics of this vapour-forming PFC model, like all previous PFC models, suffers from the fact that density and concentration evolve as separate conserved fields. These dynamics are incorrect when large bulk density changes are allowed, which is the case in solidification processes.   

In this paper we address the issue of consistently describing the density-pressure-temperature-concentration ($\rho$-$p$-$T$-$c$) space of complex binary alloys and their dynamics in the paradigm of the the XPFC alloy model of Smith \textit{et al.}~\cite{smith2017}. We begin by reinterpreting the fields of the XPFC free energy. We then generalize some of the ideas of Kocher \textit{et al.} to control density-pressure-temperature-concentration space of a binary system that can form different crystal phases. Moreover, by connecting solute concentration ($c$) and total density ($\rho$) to the individual species densities of a binary system $\rho_A$, $\rho_B$, we introduce alloy dynamics for total density $\rho$ and concentration $c$ that is consistent with the conservation of individual species densities.  

The remainder of the paper is organized as follows. Section~\ref{sec:OldModel} reviews the derivation of the original XPFC alloy model of Greenwood \textit{et al.}~\cite{greenwood2011binary}, highlighting recent improvements regarding the enthalpy of mixing. In Sec.~\ref{sec:ConsistentDynamics}, we re-introduce the definition of concentration ($c$) in the XPFC alloy model such as to be consistent with the notion of a smooth $c$ field that is coupled to a microscopic density. The dynamics of the alloy XPFC model are then re-formulated in terms of individual species densities of the binary alloy. This section also introduces algorithms for implementing pressure and volume changes during dynamical simulations. Sec.~\ref{sec:EquilibriumProperties} then studies the equilibrium properties of the reformulated XPFC alloy model in density($\rho$)-pressure($p$)-concentration($c$)-temperature($T$) space. The demonstration of the model in effecting  pressure controlled kinematics is shown in Sec.~\ref{sec:Kinematics}, while the application of the model to solute drag is located in Sec.~\ref{sec:SoluteDrag}. We conclude in Sec.~\ref{sec:Conclusion}.

\section{\label{sec:OldModel} XPFC Model of Binary Alloys}

The binary XPFC alloy model has been established for several years as a PFC  phenomenology for simulating substitutional binary alloys that crystallize into a wide range of crystalline symmetries alloys~\cite{elder2007,elder2011,lu2015,greenwood2012,fallah2012,fallah2013simulation,fallah2013atomistic,elder2010,lu2016,seymour2016,alster2017}. As with any PFC theory, it can be derived from classical density-functional theory of mixtures, where the density fields of the two components $\rho_A$ and $\rho_B$ are transformed according to the following relations
%%%%%%
\begin{equation}
\begin{split}
n =& \frac{\rho_A + \rho_B - \rho_0}{\rho_0} \\
c =& \frac{\rho_B}{\rho_A + \rho_B}
\end{split}
\end{equation}
%%%%%%%
where $n$ is the total mass density, $c$ is the concentration of the $B$ component, and $\rho_0 = \rho_A^0 + \rho_B^0$ is the total reference density around which the theory is nominally expanded.
A key assumption of the XPFC alloy theory is to assume that the $c$ field is locally smooth compared to $n$ in order to arrive at the free energy in Eq.~(\ref{XPFC_binary_Free1}). This makes it possible to couple the microscopic density field $n$ to only the long wavelength properties of concentration $c$, as tacitly reflected in form of the XPFC alloy model reviewed below.

With the variable transforms and assumptions stated above, the free energy functional for an XPFC binary alloy becomes 
%%%%%%%%%%%%
\begin{equation}
\begin{split}
\frac{\Delta F}{\rho_0k_{\rm B}T} = \int{\rm d}^3r\Big\{&\frac{n}{2}(1-C_{nn}*)n - \frac{t}{6}n^3 + \frac{v}{12}n^4 \\ & 
+ w (n+1)
S_\mathrm{mix}
- \frac{1}{2}c\;C_{cc}*c\Big\},
\end{split}
\label{XPFC_binary_Free1}
\end{equation}
%%%%%%%%%%%%%
where we have introduced the notation $A*B$ to represent a convolution operation, while $t$, $v$, and $w$ are phenomenological fitting parameters, and $S_\mathrm{mix}$ is the entropy of mixing, given by
%%%%%%
\begin{equation}
    S_\mathrm{mix}=c\ln\left(\frac{c}{c_0}\right)+(1-c)\ln\left(\frac{1-c}{1-c_0}\right),
    \label{XPFC_alloy_mix}
\end{equation}
%%%%%%%%
where $c_0$ is a reference concentration $\rho_B^0 / \rho_0$. Ordering in the total mass density are controlled through the direct correlation function $C_{nn}$,  which has the form
%%%%%%%%
\begin{equation}
    C_{nn}(|r-r^\prime|) = \sum_{i\in\mathcal{N}}\xi_i(c)C_{i}(|r-r^\prime|),
    \label{XPFC_binary_Ceff}
\end{equation}
%%%%%%%%%%%
where $\mathcal{N}$ sums over an arbitrary number of ordered phases possible and $C_i$ encodes for the density ordering of the phase $i$. Equation~(\ref{XPFC_binary_Ceff}) allows for changes in crystal structure between phases by coupling each $C_i$ to the local composition $c$, via the functions $\xi_i(c)$. 
One such interpolation function for eutectic alloys was proposed by Greenwood \textit{et al.}~\cite{greenwood2011binary} as
\begin{equation}
\begin{split}
    \xi_A(c) &= 1 - 3c^2 + 2c^3 \, ,\\
    \xi_B(c) &= \xi_A(1-c) \, ,  
\end{split}
\end{equation}
and will be used throughout the remainder of the paper paper. 

In XPFC models, the short wavelength properties of the correlation functions $C_i$ are controlled in Fourier space by superimposing a series of Gaussian peaks of the form
%%%%%%%%%
\begin{equation}
    C_{i}(k) = \overline{\sum_{i}} e^{-\frac{T}{T_M}}e^{-\frac{(k-k_i)^2}{2\sigma_i^2}},
    \label{individual_XPFC_corr}
\end{equation}
%%%%%%%%
where the bar over the sum denotes taking the maximal value of the Gaussians wherever there is overlap, $T_M$ is the melting temperature, $k_i = 2\pi / \lambda_i$ where $\lambda_i$ is the wavelength of the given crystallographic plane, and $\sigma_i$ is the width of the Gaussian, which is related to the elasticity of the crystal. To the sum of Eq.~(\ref{individual_XPFC_corr}) can also be added a negative $k=0$ peak that can further be used to control the compressibility of emerging solid phases. 

In the model of Eq~(\ref{XPFC_binary_Free1}), correlations in concentration fluctuations are controlled through $C_{cc}(|r-r'|)$. In the long-wavelength limit, Smith \textit{et al.}~\cite{smith2017} proposed a form for $C_{cc}$ given by
%%%%%
\begin{equation}
    C_{cc}(r-r^\prime) = -w \epsilon(T)\delta(r-r^\prime) - W_c\nabla^2\delta(r-r^\prime) \,.
    \label{CC_eff_Smith}
\end{equation}
%%%%%
Applying the divergence theorem, under suitable boundary conditions, to the $C_{cc}$ term in Eq.~(\ref{XPFC_binary_Free1}) yields a more familiar form of the XPFC binary alloy free energy functional,
%%%%%%%%%%%%
\begin{equation}
\begin{split}
\frac{\Delta F}{\rho_0k_{\rm B}T} = \int{\rm d}^3r\Big\{&\frac{n}{2}(1-C_{nn}*)n - \frac{t}{6}n^3 + \frac{v}{12}n^4 \\ & + w(n+1)
S_\mathrm{mix}
+ \frac{1}{2}W_c|\nabla c|^2
\\
&+\frac{1}{2}w\epsilon(T)(c-c_0)^2\Big\}.
\end{split}
\label{simplified_XPFC_binary_Free1}
\end{equation}
%%%%%%%%%%%%%%%
Equation~(\ref{simplified_XPFC_binary_Free1}) contains energy penalties for concentration gradients as prescribed by Cahn and Hilliard~\cite{cahn1958}, as well as the leading order enthalpy of mixing contribution.

\section{\label{sec:ConsistentDynamics} Improvements to the consistency of the XPFC Alloy Model}

The free energy as derived in Sec.~\ref{sec:OldModel}, and its dynamics, have been shown to qualitatively reproduce a host of physical phenomena~\cite{elder2007,elder2011,lu2015,greenwood2012,fallah2012,fallah2013simulation,fallah2013atomistic,elder2010,lu2016,seymour2016,alster2017}. The model has been especially convenient for calculating equilibrium properties of systems, since $n$ and $c$ are the natural variables used in physical metallurgy. However, when considering dynamical process, particularly those accompanied by significant density changes, difficulties arise from the use of these variable due to the fact $c$ is not truly a conserved field and as such is not governed by conserved  Langevin dynamics. Strictly speaking, the conserved variables of our system are $\rho_A$ and $\rho_B$. Another problem with the XPFC alloy model is that it tacitly assumes that the concentration $c$ field is smooth, although small spatial oscillations do in fact develop in it in some processes. These issues are both addressed next.

\subsection{Re-Defining XPFC Concentration}
\label{redefinng_c}

In order to derive dynamics for this system in the governing density variables of $\rho_A$ and $\rho_B$ of components $A$ and $B$, respectively, we revisit the underlying assumption made in the original derivation of Greenwood \textit{et al}~\cite{greenwood2011binary}. Therein, a critical explicit assumption was the smoothness of the $c$-field, which implies the substitutional nature of the alloy model. In doing so, it allowed the simplification of convolution integrals involving $c$. These assumptions, and consequent manipulations of the free energy ensured the impossibility of a return to a $\rho_{A}$--$\rho_{B}$ formulation due to the information lost in the smoothing process. To rectify this, while maintaining a connection with the variables $c$ and $n$, we consider the relationship that already exists between $n$ and $c$ and $\rho_A$ and $\rho_B$. Namely, we insist on the smoothness of the concentration field by positing a relationship of the form 
%%%%%%%%
\begin{equation}
c\equiv \frac{\chi * \rho_B}{\chi * (\rho_A + \rho_B)}
\label{eq:SmoothedConcentration}
\end{equation} 
%%%%%%%
where $\chi$ is a smoothing kernel that retains long wavelength information of the field upon which it operates. This is similar to the use of smoothing kernels employed by Kocher \textit{et al.}~\cite{kocher2015} and is defined in reciprocal space as
%%%%%%%%%%
\begin{equation}
\tilde{\chi}(k) \equiv e^{-\frac{k^2}{2\lambda_c}},
\label{smoothing_kernel}
\end{equation}
%%%%%%%
where $\lambda_c$ sets the cutoff wavelength. Moreover, this definition of $c$ also recovers the equilibrium definition of concentration as a bulk quantity.

We make the further improvement of scaling the free energy by a reference temperature instead of the model temperature, as was done by Kocher \textit{et al.}~\cite{kocher2015}. This introduces a factor of reduced temperature $\tau = T / T_0$ in the free energy functional, according to 
%%%%%%%%%%%%
\begin{equation}
\begin{split}
\frac{F}{\rho_0k_{\rm B}T_0} = \tau\int{\rm d}^3r\Big\{&\frac{n}{2}(1-C_{nn}*)n - \frac{t}{6}n^3 + \frac{v}{12}n^4 \\ & + w(n+1)
S_\mathrm{mix}
+ \frac{1}{2}w\epsilon(T)(c-c_0)^2
\\
&+\frac{1}{2}W_c|\nabla c|^2\Big\} + \tau\bar{F}(\rho_0,c_0),
\end{split}
\label{eq:FullFreeEnergy}
\end{equation}
%%%%%%%%%%%%%%
where Eq.~(\ref{eq:FullFreeEnergy}) also retains the free energy of the reference fluid, $\bar{F}(\rho_0,c_0)$. This term will be largely neglected throughout this paper as it will not affect either the phase diagram or the dynamics; its primary importance is in quantifying the reference pressure of our system.

\subsection{\label{subsec:Dynamics} Model Dynamics}

With the reformulation of the concentration introduced in Sec.~\ref{redefinng_c}, the dynamical evolution of the alloy system may be calculated with respect to its governing variables $\rho_A$ and $\rho_B$. Namely, conserved dynamics in these fields follow
%%%%%%%%
\begin{equation}
    \frac{\partial \rho_i}{\partial t} = \nabla\cdot(M_i\nabla\mu_i) \approx M_i\nabla^2\left(\frac{\delta F}{\delta \rho_i}\right) \, ,
    \label{eq:Dynamics}
\end{equation}
%%%%%%%%%%%%%
where $M_i$ is the mobility of the constituent density, assumed for simplicity to be a constant here, and $\mu_i$ is the chemical potential of the constituent. The chemical potential is calculated from the $(n,c)$-based free energy by means of the functional chain rule
%%%%%%%%%%%%%
\begin{equation}
\begin{split}
    \frac{\delta F[n,c](r^\prime)}{\delta\rho_i(r)} = \int{\rm d}^3r^{\prime\prime}\biggr[&\frac{\delta n(r^{\prime\prime})}{\delta\rho_i(r)}\frac{\delta F[n,c](r^\prime)}{\delta n(r^{\prime\prime})} \\&+ \frac{\delta c(r^{\prime\prime})}{\delta\rho_i(r)}\frac{\delta F[n,c](r^\prime)}{\delta c(r^{\prime\prime})}\biggr]\,.
\end{split}
\end{equation}
%%%%%%%%%%%%%%%
Evaluation of this chain rule gives the respective chemical potentials of each species as 
%%%%%%%%%%%%%%%%%%%%%%%%%
\begin{widetext}
\begin{align}
\frac{\mu_A}{\rho_0 k_\mathrm{B}T_0} = \begin{aligned}[t] &
\tau \left\{(1-C_{nn})*n(r) - \frac{t}{2}n^2(r)+\frac{v}{3}n^3(r) + w\left(c(r)\ln\left(\frac{c(r)}{c_0}\right) + (1-c(r))\ln\left(\frac{1-c(r)}{1-c_0}\right)\right)\right\}\\&
-\tau \int{\rm d}^3r^{\prime\prime}\chi(r^{\prime\prime}-r)\frac{c(r^{\prime\prime})}{\int{\rm d}^3r^{\prime\prime\prime}\chi(r^{\prime\prime\prime}-r^{\prime\prime})(n(r^{\prime\prime\prime})+1)}\times
\begin{aligned}[t]
\bigg[&w(n(r^{\prime\prime})+1)\left(\ln\left(\frac{c(r^{\prime\prime})}{c_0}\right) - \ln\left(\frac{1-c(r^{\prime\prime})}{1-c_0}\right)\right)\\&
+w\epsilon(T)(c(r^{\prime\prime})-c_0)-W_c\nabla^2c(r^{\prime\prime})\bigg]
\end{aligned}\\&
+\frac{\tau}{2}\int{\rm d}^3r^{\prime\prime}\chi(r^{\prime\prime}-r)\frac{c(r^{\prime\prime})}{\int{\rm d}^3r^{\prime\prime\prime}\chi(r^{\prime\prime\prime}-r^{\prime\prime})(n(r^{\prime\prime\prime})+1)}n(r^{\prime\prime})\int{\rm d}^3r^\prime n(r^\prime)\frac{\partial C_{nn}(r^\prime,r^{\prime\prime})}{\partial c} \, ,
\label{eq:ChemicalPotentialA}
\end{aligned}
\end{align}
\begin{align}
\frac{\mu_B}{\rho_0 k_\mathrm{B}T_0} = \begin{aligned}[t] &
\tau \left\{(1-C_{nn})*n(r) - \frac{t}{2}n^2(r)+\frac{v}{3}n^3(r) + w\left(c(r)\ln\left(\frac{c(r)}{c_0}\right) + (1-c(r))\ln\left(\frac{1-c(r)}{1-c_0}\right)\right)\right\}\\&
+\tau \int{\rm d}^3r^{\prime\prime}\chi(r^{\prime\prime}-r)\frac{1-c(r^{\prime\prime})}{\int{\rm d}^3r^{\prime\prime\prime}\chi(r^{\prime\prime\prime}-r^{\prime\prime})(n(r^{\prime\prime\prime})+1)}\times
\begin{aligned}[t]
\bigg[&w(n(r^{\prime\prime})+1)\left(\ln\left(\frac{c(r^{\prime\prime})}{c_0}\right) - \ln\left(\frac{1-c(r^{\prime\prime})}{1-c_0}\right)\right)\\&
+w\epsilon(T)(c(r^{\prime\prime})-c_0)-W_c\nabla^2c(r^{\prime\prime})\bigg]
\end{aligned}\\&
-\frac{\tau}{2}\int{\rm d}^3r^{\prime\prime}\chi(r^{\prime\prime}-r)\frac{1-c(r^{\prime\prime})}{\int{\rm d}^3r^{\prime\prime\prime}\chi(r^{\prime\prime\prime}-r^{\prime\prime})(n(r^{\prime\prime\prime})+1)}n(r^{\prime\prime})\int{\rm d}^3r^\prime n(r^\prime)\frac{\partial C_{nn}(r^\prime,r^{\prime\prime})}{\partial c} \, .
\label{eq:ChemicalPotentialB}
\end{aligned}
\end{align}
\end{widetext}
%%%%%%%%%%%%%%%%%%%%%%

We draw particular attention here to the terms premultiplying the interpotential $\mu_\mathrm{inter}=\delta F / \delta c$, which demonstrate why Eq.~(\ref{eq:SmoothedConcentration}) must be written with smoothing kernels in both the numerator and denominator. If a single smoothing kernel is used which acts on the traditional definition of the concentration, this prefactor will have a term which goes as $1 / (n(r) + 1)$ which can result in numerical instabilities due to the oscillations in the $n$ field.

\subsection{\label{subsec:PressureControl} Dynamical Pressure Control and Volume Dynamics}

To describe pressure changes during dynamical simulations, we follow the work of Kocher \textit{et al}~\cite{kocher2015} and utilize the grand potential density as an approximation for system pressure, thus allowing for isobaric or pressure controlled systems. The grand potential density is defined as
%%%%%%%%%%%%%%%
\begin{equation}
    \omega = \frac{1}{V}\int_V{\mathrm d}^3r\left(f - \sum_i \mu_i\rho_i\right),
    \label{grand_potential_den}
\end{equation}
%%%%%%%%%%%%%%%%%%%
where $f$ is the free energy density of the multi-component alloy. Specializing to a binary alloy, the grand potential density can now be calculated using Eqs.~(\ref{eq:FullFreeEnergy}),~(\ref{eq:ChemicalPotentialA}), and~(\ref{eq:ChemicalPotentialB}). Generalization to a multi-component alloy is straightforward.

For single component systems Ref.~\cite{kocher2015} used a two step process. The first step is a mass conserving flux of the form
%%%%%%%%%%%%%%
\begin{equation}
    J_n = -n_0(t) + n_0(t_0)\left(\frac{\Delta x(t_0)}{\Delta x(t)}\right)^{d},
    \label{eq:GabyDensityFlux}
\end{equation}
%%%%%%%%%%%%%%%%%%%%%%%
where $\Delta x$ is the side length of the volume element, which is assumed to be a $d$-cube where $d$ is the dimensionality of the system, $t$ is the simulation time, $t_0$ is the initial time and $n_0$ is the global average density of the system.
This flux is added to each volume element. The second step is changing the volume element with a proportional feedback loop defined as
%%%%%%%%%%%%%%%%
\begin{equation}
    \Delta x(t) = \Delta x(t-\Delta t) + \frac{\Delta t M_p}{d(\Delta x)^{d-1}}(\omega - P_0) \, .
    \label{eq:GabyVolumeChange}
\end{equation}
%%%%%%%%%%%%%%%%%%%%%
where $P_0$ is the target pressure, $\Delta t$ is the length of a numerical time-step, and $M_p$ is the mobility parameter for readjustments to the volume element --- \textit{i.e.} the ability of the system to respond to pressure differentials.

\begin{figure}
    \centering
    \includegraphics[width=\linewidth]{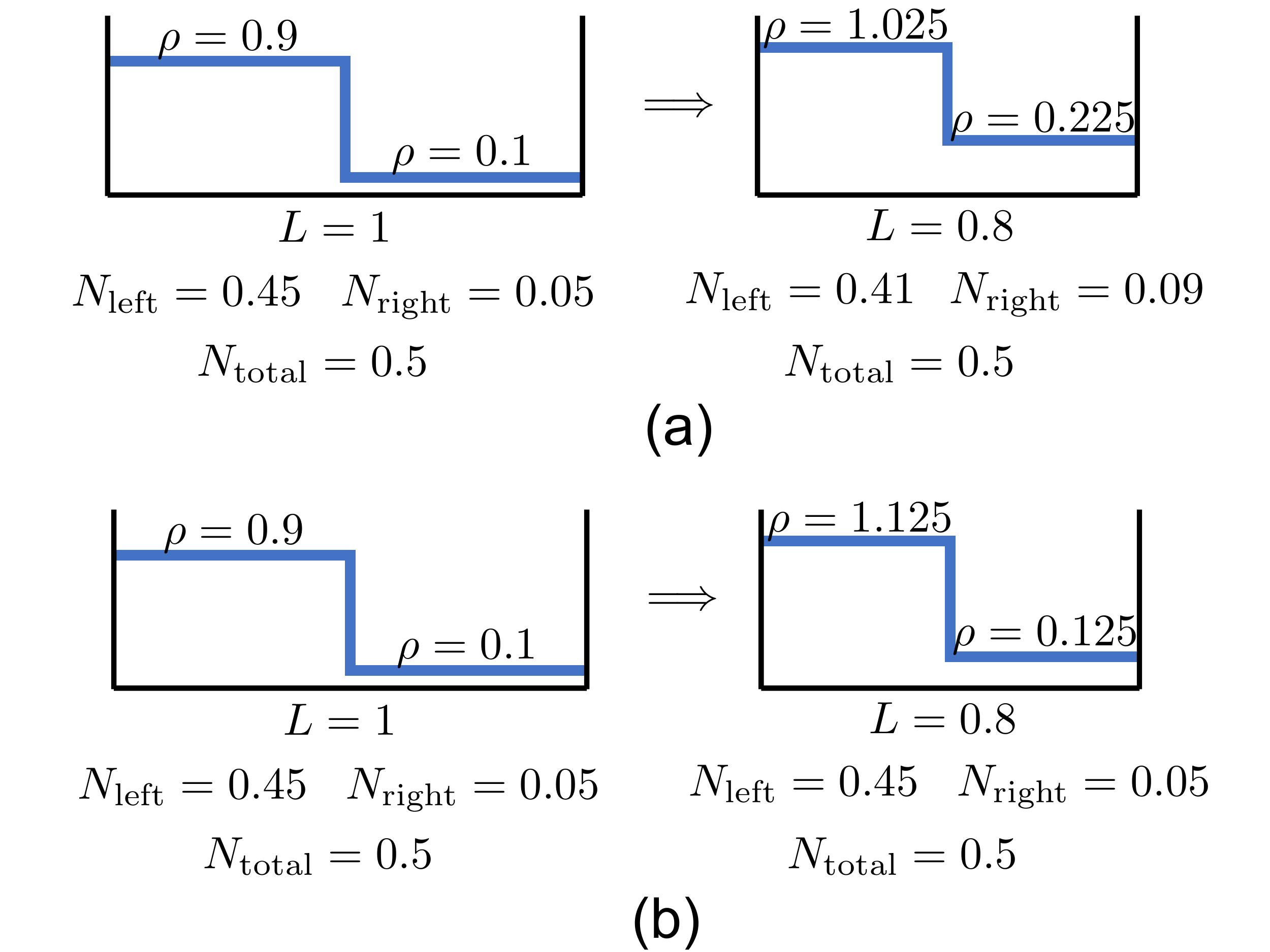}
    \caption{A two phase system undergoing a number conserving change in volume as described by (a) Eqs.~(\ref{eq:GabyDensityFlux}) and (\ref{eq:GabyVolumeChange}) and (b) Eqs.~(\ref{eq:NewDensityFlux}) and (\ref{eq:NewVolumeChange}).}
    \label{fig:DiffusionThoughtExp}
\end{figure}

While such a density flux and feedback loop conserves the total mass of the system it violates the physics of diffusion transport. Under a compression of a volume element, a low density element and a high density element both receive the same density flux into the system. Local mass is thus not conserved as the low density volume element has gained a larger fraction of its density than the high density element; mass has effectively moved instantaneously from the high density region to the low density region. We illustrate a one-dimensional example of this phenomenon in Fig.~\ref{fig:DiffusionThoughtExp}(a), and the correction using the scheme that we implement in this work (discussed below) to remedy this problem in Fig.~\ref{fig:DiffusionThoughtExp}(b). 
%%%%%%%%%%%%%
%%%%%%%%%%%%%

To maintain accurate diffusion behaviour we propose a slightly different mass conserving step. Conservation of total number ($N$), or mass, requires that
%%%%%%%%%%%%%%
\begin{equation}
    \frac{\partial N}{\partial t} = \frac{\partial(\bar{\rho} V)}{\partial t} = 0 \, ,
    \label{eq:massConservation}
\end{equation}
%%%%%%%%%%%%%%%
where $V$ is the volume and $\bar{\rho}$ is the average density of the system or sub-system. After expressing the volume as a numerical grid, $V=\Pi_{i=1}^d \Delta x N_i$, and combining with the volume feedback loop proposed by Kocher \textit{et al.}, 
%%%%%%%%%%%%%%%
\begin{equation}
    \frac{\partial \Delta x}{\partial t} = \frac{M_p}{d \, (\Delta x)^{d-1}}(\omega - P_0) \, ,
\end{equation}
%%%%%%%%%%%%%%%
Eq.~\ref{eq:massConservation} can be rewritten as
%%%%%%%%%%%%%%
\begin{equation}
    \frac{\partial \bar{\rho}}{\partial t} = - \frac{M_p(\omega - P_0)}{(\Delta x)^d}\bar{\rho} \, .
    \label{eq:ConservedN}
\end{equation}
%%%%%%%%%%%%%%%
In pursuit of a closed form solution for the numerical update scheme of Eq.~(\ref{eq:ConservedN}), we posit that this equation can be solved via separation of variables and integration,
%%%%%%%%%%%%%%%%%
\begin{equation}
    \int\frac{\mathrm{d}\bar{\rho}}{\bar{\rho}} = \int\mathrm{d}t\frac{M_p(\omega-P_0)}{(\Delta x)^d} \, ,
\end{equation}
%%%%%%%%%%%%%%%%%
which leads to the numerically approximated solution
%%%%%%%%%%%%%
\begin{equation}
    \bar{\rho}(x,t+\Delta t) = \bar{\rho}(x,t)e^{-\frac{M_p(\omega-P_0)\Delta t}{(\Delta x)^d}} \, .
    \label{eq:NewDensityFlux}
\end{equation}
%%%%%%%%%%%%%%%%
After a point-wise evaluation of this local density update on each grid element, we then ensure total conservation of mass by appropriately updating $\Delta x$. This update leads to a revision of
Eq.~(\ref{eq:GabyVolumeChange}) of the form, 
%%%%%%%%%%%%%
\begin{equation}
    J_{\Delta x} = -\Delta x(t) + \Delta x(t_0)\left(\frac{\int_{V_\mathrm{total}}\mathrm{d}^dr\rho(t)}{\int_{V_\mathrm{total}}\mathrm{d}^dr\rho(t_0)}\right)^{1/d} \, .
    \label{eq:NewVolumeChange}
\end{equation}
%%%%%%%%%%%%%
Due to the numerical constraints of the FFTW~\cite{FFTW} library used in our dynamics, the volume element must be a constant throughout the simulation domain. Thus, we must apply Eq.~(\ref{eq:NewVolumeChange}) to each volume element individually in such a way that the global update of [common] volume element size conserves mass in local sub-domains of of volume elements in the system. As any arbitrary sub-domain of the system must conserve its own local mass we choose to update the smallest possible sub-domain --- \textit{i.e.} each pixel --- as it is both numerically simplest and ensures that any larger sub-domain will also be conserved.

Due to the exponential nature of Eq.~(\ref{eq:NewDensityFlux}), it is imperative that the argument of the exponential remain reasonably small. To ensure this smallness, we utilize a ``guess--check-correct" algorithm as follows: 
\begin{enumerate}
    \item We calculate the expected change in volume element, $\delta_\mathrm{guess}\Delta x$ using the original formulation given by equation~(\ref{eq:GabyVolumeChange}) using the largest allowable time-step $\Delta t = \Delta t_\mathrm{max}$.
    \item This value is compared to a cutoff value $\delta_\mathrm{max}\Delta x$.
    \item If and only if the expected change is larger than the cutoff we scale the time-step as follows: $\Delta t = \Delta t_\mathrm{max}\frac{\delta_\mathrm{max}\Delta x}{\delta_\mathrm{guess}\Delta x}$.
    \item Eqs.~(\ref{eq:NewDensityFlux}) and (\ref{eq:NewVolumeChange}) are applied.
\end{enumerate}

\section{\label{sec:EquilibriumProperties} Equilibrium Properties of Model}

The equilibrium properties of a binary alloy, where density variations are considered, are defined in terms of the Helmholtz free energy by the system of equations~\cite{landau1980,jugdutt2015}. 
%%%%%%%%%%%%%%%%%%%%
\begin{align} \nonumber
\frac{1}{\rho_{s}} \frac{\partial F_{s}\left(c_{s}, \rho_{s}\right)}{\partial c_{s}} &=\mu^{\mathrm{eq}} \, , \\
\nonumber \frac{1}{\rho_{L}} \frac{\partial F_{L}\left(c_{L}, \rho_{L}\right)}{\partial c_{L}} &=\mu^{\mathrm{eq}} \, , \\
\rho_{s} \frac{\partial F_{s}\left(c_{s}, \rho_{s}\right)}{\partial \rho_{s}}-F_{s}\left(c_{s}, \rho_{s}\right) &=p \, , \\ \nonumber
\rho_{L} \frac{\partial F_{L}\left(c_{L}, \rho_{L}\right)}{\partial \rho_{L}}-F_{L}\left(c_{L}, \rho_{L}\right) &=p \, , 
\\ \nonumber
\frac{F_{L}\left(c_{L}, \rho_{L}\right)}{\rho_{L}}-\frac{F_{s}\left(c_{s}, \rho_{s}\right)}{\rho_{s}} =\left(c_{L}-c_{s}\right)& \mu^{\mathrm{eq}}-\left( \tfrac{1}{\rho_{L}}-\tfrac{1}{\rho_{s}}\right) p \, .
\end{align}
%%%%%%%%%%%%%%%
where we have used the short form notation $F_i$, $\rho_i$, and $c_i$ to denote the free energy density, total density, and concentration of phase $i$, respectively, where $i=L$ denotes a liquid and $i=s$ a solid phase. Here, $\mu$ is the chemical potential and $p$ the pressure. 

The solution to the above system of equations in terms of the free energy defined in Sec.~\ref{sec:OldModel} is highly non-trivial~\cite{jugdutt2015}. However, the system of equations can be significantly simplified by a change of variables such that
%%%%%%%%%%%%%%%
\begin{equation}
    \begin{split}
        \mathcal{F} = \nu F\;,
    \end{split}
\end{equation}
%%%%%%%%%%%%%%%%%
where $\nu = 1 / \rho$ is the molar volume. The equilibrium equations then become
%%%%%%%%%%%%%%
\begin{align} \nonumber
\frac{\partial \mathcal{F}_{s}\left(c_{s}, \nu_{s}\right)}{\partial c_{s}} &=\mu^{\mathrm{eq}} \\ \nonumber
\frac{\partial \mathcal{F}_{L}\left(c_{L}, \nu_{L}\right)}{\partial c_{L}} &=\mu^{\mathrm{eq}} \\
\frac{\partial \mathcal{F}_{s}\left(c_{s}, \nu_{s}\right)}{\partial \nu_{s}} &=-p \label{eq:SimplifiedEquilibrium}\\ \nonumber
\frac{\partial \mathcal{F}_{L}\left(c_{L}, \nu_{L}\right)}{\partial \nu_{L}}&=-p \\ \nonumber
\mathcal{F}_{L}\left(c_{L}, \nu_{L}\right)-\mathcal{F}_{s}\left(c_{s}, \nu_{s}\right) &=\left(c_{L}-c_{s}\right) \mu^{\mathrm{eq}}-\left(\nu_{L}-\nu_{s}\right) p
\end{align}
%%%%%%%%%%%%%%%%
Eqs.~(\ref{eq:SimplifiedEquilibrium}) define the equations of a common tangent plane in ($\nu$-$c$-$T$) space. In the traditional formulations of binary PFC models --- with the exception of one amplitude model~\cite{jugdutt2015} --- the density (molar volume) are assumed to be constant across the different phases in equilibrium, which reduces Eqs.~(\ref{eq:SimplifiedEquilibrium}) to
%%%%%%%%%%%%%%%%%%
\begin{align} \nonumber
\frac{\partial F_{s}\left(c_{s}, \rho\right)}{\partial c_{s}} &=\mu^{\mathrm{eq}} \, , \\
\nonumber \frac{\partial F_{L}\left(c_{L}, \rho\right)}{\partial c_{L}} &=\mu^{\mathrm{eq}} \, , \\ \nonumber  F_{L}\left(c_{L}, \rho\right)-F_{s}\left(c_{s}, \rho\right) &=\left(c_{L}-c_{s}\right) \mu^{\mathrm{eq}} \, ,
\end{align}
%%%%%%%%%%%%%%%%%%
which define the equations of a common tangent line in ($c$-$T$) space. 

We use a similar, but distinct, approach to define the isobaric phase diagram from from Eqs.~(\ref{eq:SimplifiedEquilibrium}) as follows:  for a fixed pressure we rearrange the last of Eqs.~(\ref{eq:SimplifiedEquilibrium}) to the form
%%%%%%%%%%%%%%
\begin{align}
   &&(\mathcal{F}_{L}\left(c_{L}, \nu_{L}\right)+\nu_{L}\,p)-(\mathcal{F}_{s}\left(c_{s}, \nu_{s}\right)+\nu_{s}\,p) \nonumber \\
   &&=\left(c_{L}-c_{s}\right) \mu^{\mathrm{eq}} \, .
   \label{eq:TangentEquation}
\end{align}
%%%%%%%%%%%%%%%%
The  first two of Eqs.~(\ref{eq:SimplifiedEquilibrium}) and Eq.~(\ref{eq:TangentEquation}) now define a common tangent in concentration along an isobaric surface, with the isobaric constraint of said surface being enforced by the third and fourth equations of Eqs.~(\ref{eq:SimplifiedEquilibrium}).

While Eqs.~\ref{eq:SimplifiedEquilibrium} are shown in dimensional form, an identical set of dimensionless equations can be constructed by scaling the variables similarly to Eq.~(\ref{eq:FullFreeEnergy}):
\begin{align}
    \bar{\mathcal{F}} &= \frac{\mathcal{F}}{ k_\mathrm{B}T_0} = \frac{\rho_0}{\rho}\frac{F}{\rho_0k_\mathrm{B}T_0}\\
    \bar{\mu}^\mathrm{eq} &= \frac{\mu^\mathrm{eq}}{k_\mathrm{B}T_0}\\
    \bar{p} &= \frac{\nu_0p}{k_\mathrm{B}T_0} = \frac{p}{\rho_0k_\mathrm{B}T_0}\\
    \bar{\nu} &= \frac{\nu}{\nu_0} = \frac{\rho_0}{\rho}
\end{align}
which allows the use of scaled PFC free energies of the form of Eq.~(\ref{eq:FullFreeEnergy}) to be used in the construction of phase diagramsm and in the subsequent dynamics.

To determine the equilibrium phase diagram it is necessary to define a free energy curve along the isobaric surface of the free energy landscape. To do so we follow the mode expansion methodology pioneered by Kirkwood and Monroe~\cite{kirkwood1941} in 1941 and later refined by Yousseff and Ramakrishnan~\cite{ramakrishnan1979} and approximate the total density and  total molar volume
of a phase as 
%%%%%%%%%%%%%%%%
\begin{equation}
    \begin{split}
        n &\approx n_0 + \sum_j A_j \sum_\alpha e^{\mathrm{i}k_{\{j,\alpha\}} \cdot r} \\
        n_0 &= \frac{1}{\bar{\nu}} - 1 \\
        % c &= c_0
    \end{split}
\end{equation}
%%%%%%%%%%%%%%%%
where $A_j$ are taken as non-zero in a solid phase. The index $j$ defines a given family of modes in a solid assumed to have the same amplitude, while $k_{\{j,\alpha\}}$ is the wave-vector of the lattice plane $\alpha$ within the family of modes $j$, and $r$ is the spatial coordinate.

Substituting the density expansion ansatz in a microscopically varying free energy density of a PFC model, and integrating out short scale variations over the scale of a crystal unit cell, leads to a mesoscale free energy representation of the system of the form $F(n_0(\nu),\{A_j\},c)$. Moreover, it is noted that for a bulk phase, the derivative 
%%%%%%%%%%%%
\begin{equation}
        \frac{\partial \bar{\mathcal{F}}\left(\bar{\nu}(n_0),\{A_j\},c\right)}{\partial \bar{\nu}}  = \omega(\bar{\nu}(n_0),\{A_j\},c),
\end{equation}
%%%%%%%%%%%%%
and thus we can enforce the pressure constraint by equating the amplitude expanded grand potential density to the negative of the target pressure. 

The protocol for evaluating the equilibrium states of the alloy proceeds next as follows: for each value of the average concentration $c$ we numerically minimize the set of amplitudes $\{n_0, A_j\}$ and wave-vector $k$ subject to the pressure constraint to build the isobaric free energy curve using Mathematica~\cite{Mathematica}. For the special case of a single crystalline structure a single minimization will yield the free energy values for both the solid, $\{A_j\}\neq 0$, and liquid $\{A_j\}=0$. In the more generic case of multiple crystalline structures, a minimization is required for each solid phase; the set of free energies are then compared and only the minimum is kept. This [minimum] free energy is then shifted as per Eq.~(\ref{eq:TangentEquation}) and passed to Mathematica's convex hull finding algorithm, which acts as a common tangent finding algorithm when passed a 1-dimensional landscape. This procedure is repeated for each pressure and/or temperature of interest.

Using the conserved dynamics described by Eqs.~(\ref{eq:Dynamics}) we validate the accuracy of the mode expansion approximation used to generate phase diagrams for the case of a eutectic alloy. Without loss of generality, we consider an alloy of structurally similar elements, differing only in the equilibrium lattice parameters of the constituent elements.

The initial conditions are set algorithmically based on a reference system to ensure that the system does not exhibit a pressure significantly different than the set target. The algorithm is as follows:
\begin{enumerate}
    \item The temperature, mean composition, expected phases, and initial phase composition offset relative to the mean are selected.
    \item The structure or structures are seeded into the liquid, and both phases are offset in composition in opposite directions from the mean by a small offset. For example, for an expected coexistence between the $\alpha$ solid and liquid, the solid would be offset by $-c_{\mathrm{offset}}$ while the liquid would be offset by $+c_{\mathrm{offset}}$. We found that this offset is not strictly necessary, but helps to speed up the equilibration so long as the offset is in the proper direction. 
    \item If a liquid-solid coexistence is selected, the solid is given a small total density increase.
    \item If a eutectic coexistence is selected then a bi-crystal is seeded with a given misorientation.
    \item Equation~(\ref{eq:NewDensityFlux}) is iterated without changing the volume element until the system is suitably
    close to the target pressure, $|\omega - P_0| \leq 10^{-6}$.
\end{enumerate}
Once these steps are complete, the total number of particles is calculated and will be conserved in all further iterations; this state serves as the initial condition for the simulation.

\begin{figure}[h]
    \centering
    \includegraphics[width=0.95\linewidth]{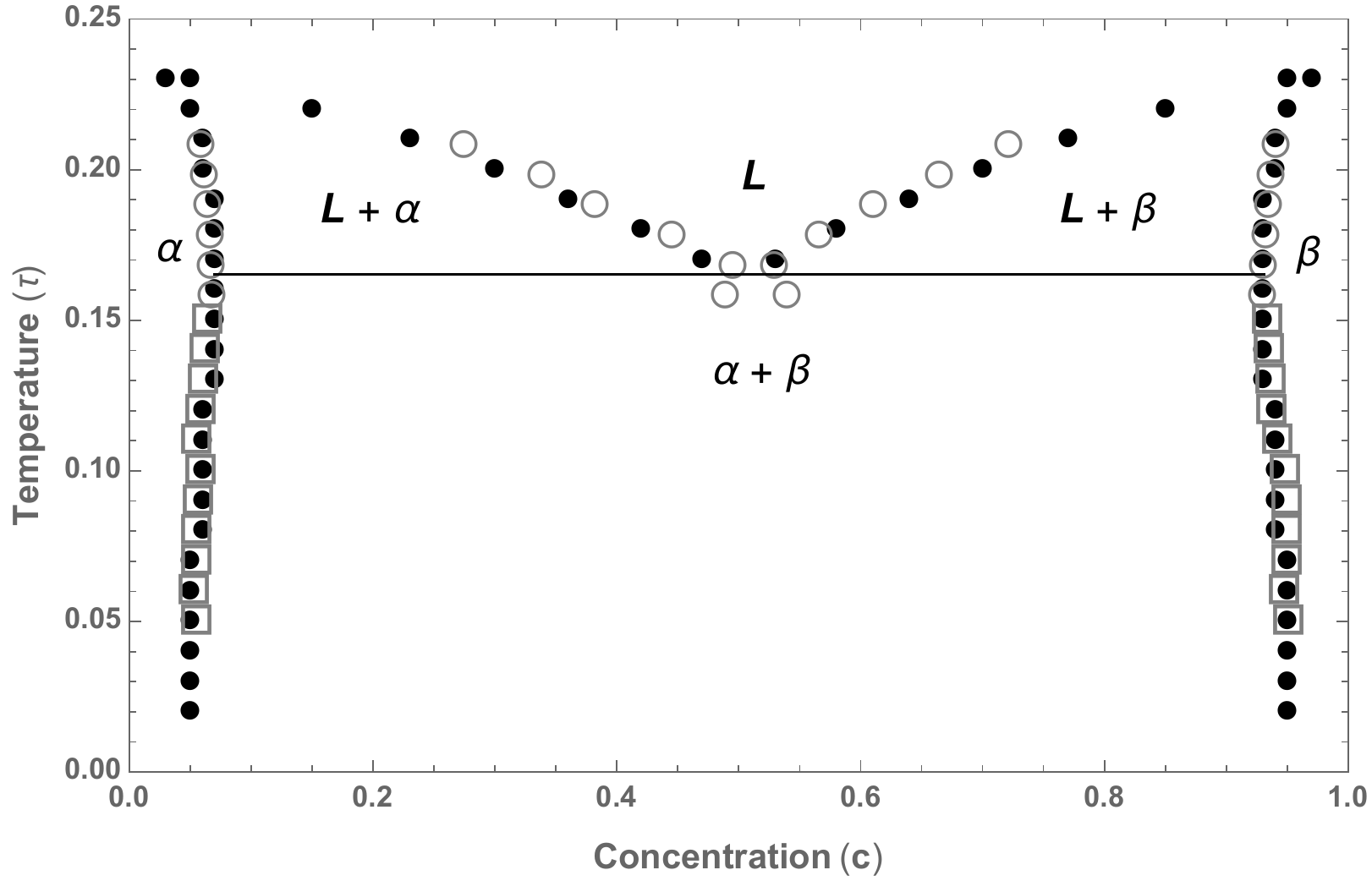}
    \caption{ Constant pressure phase diagram for a eutectic triangle-triangle system. The reduced model pressure of $P_0=0.01$. The parameters used in the ideal free energy were $t = 1.4$, $v = 1.0$, $w=0.02$, $c_0=0.5$, and $\epsilon(T) = 0$ for simplicity. The correlation kernel of the excess free energy used parameters $k^{(A)}_{10} = 2\pi$, $\sigma^{(A)}_{10} = 0.8$, $T^{(A)}_M = 1.0$, $k^{(B)}_{10} = 7.83185307179586 \approx 2.5\pi$, $\sigma^{(B)}_{10} = 0.8$, and $T^{(B)}_M =1.0$.
    The equilibrium coexistence concentrations as numerically calculated by mode expansion are shown in filled black circles while the concentrations extracted from the bulk concentrations of the dynamical simulations between liquid-solid coexistence are shown in hollow gray circles and those values extracted from dynamical simulations for solid-solid coexistence are shown in hollow gray squares.}
    \label{fig:PhaseDiagramCompare}
\end{figure}

All simulations used to verify our phase diagram were performed on a 128 by 1024 pixel grid using a square volume element with initial size $\Delta x(t=0)\approx0.0815$ and reference temperatures of $T_0=T_M=1.0$. To assist with rapid equilibration, the mobilities were set to be extremely high compared to the pressure relaxation coefficient: $M_A=M_B=20$, $M_p=0.1$. After equilibration of the simulation the resulting bulk concentration and bulk average total density for each phase is compared to that of the numerically approximated phase diagram as shown in Fig.~\ref{fig:PhaseDiagramCompare}. To avoid repeated overhead in the form of initialization, wherever possible the simulations were continued and quenched by steps of $\Delta T = 0.01$ over 10000 simulation steps. For the eutectic simulations, a misorientation of 0.1 rad $\approx 5.73^\circ$ was used. When deep in the eutectic coexistence region, the composition profiles of a bi-crystal exhibit long wavelength oscillations about their equilibrium values on length-scales greater than the lattice planar wavelengths. This is due to the fact that for a solid-solid system with two differing lattice constants it is not possible to create a box size commensurate to zero stress in both crystals.

\section{\label{sec:Kinematics} Pressure Controlled Phase Transformation Kinetics}

In this section we demonstrate the ability to control various non-equilibrium phase transformations through the system pressure. Each of the following demonstrations can be achieved through quenches in temperature, however, here, we will be considering an isothermal system subjected to compression or tension only.

In each of the following simulations we will consider a $128 \times 1024$ pixel grid with square volume elements with initial size $\Delta x \approx 0.0818$ and a constant reduced temperature $\tau=0.15$, with reference temperatures of $T_0=T_M=1.0$. The ideal free energy fitting parameters are $t=1.4$, $v=1.0$, $w=0.02$, and $c_0=0.5$. For simplicity we will set the enthalpy of mixing $\epsilon(T)=0$ for all temperatures. We consider the two-phase equilibrium system consisting of $\alpha$ and $\beta$ solid phases, with each differing in their lattice parameters. To assist with rapid equilibration, the mobilities were set to be extremely high compared to the pressure relaxation coefficient: $M_A=M_B=20$, $M_p=0.1$.

As a point of clarification, in the following sections we will often refer to processes occurring over a fixed number of ``simulation steps'', which refer to a scaled physical time rather than the actual numerical time-steps due to the adaptive
nature of the dynamical time-stepping discussed in Sec.~\ref{subsec:PressureControl}.
The times have all been rounded to the nearest output call.

\subsection{\label{subsec:Premelting} Control of Pre-melted Inter-phase Boundaries}

When solids containing interfaces or interphase boundaries approach their melting temperature, they have a propensity to undergo a phenomenon known as pre-melting~\cite{hsieh1989,adland2013,straumal2014}. During this transition, a disordered, meta-stable liquid-like film forms between the abutting solid phases. With the width of the liquid layer depending on the energetic differences between bulk solid and liquid phases, and thus decreases with decreasing temperature. The formation of the liquid layer is linked to the cost of maintaining the inter- face or phase boundary energy or its decomposition to distinct solid-liquid interfacial energy. For the case of our eutectic system, below the eutectic, the condition for pre-melting exists when  $\gamma_{\alpha\beta} > \gamma_{\alpha L}+\gamma_{\beta L}$.

To avoid the growth of either phase this simulation is set up with a system concentration of  $\langle c\rangle = 0.5$ such that we have equal phase fractions. 
The system parameters are the same as those described in the caption of Fig.~\ref{fig:PhaseDiagramCompare}. The misorientation between grains in the bi-crystal is initially set to 0.2 rad $\approx 11.459^\circ$
%%%%%%%%%%%%%%
\begin{figure}
    \centering
    \includegraphics[width=0.95\linewidth]{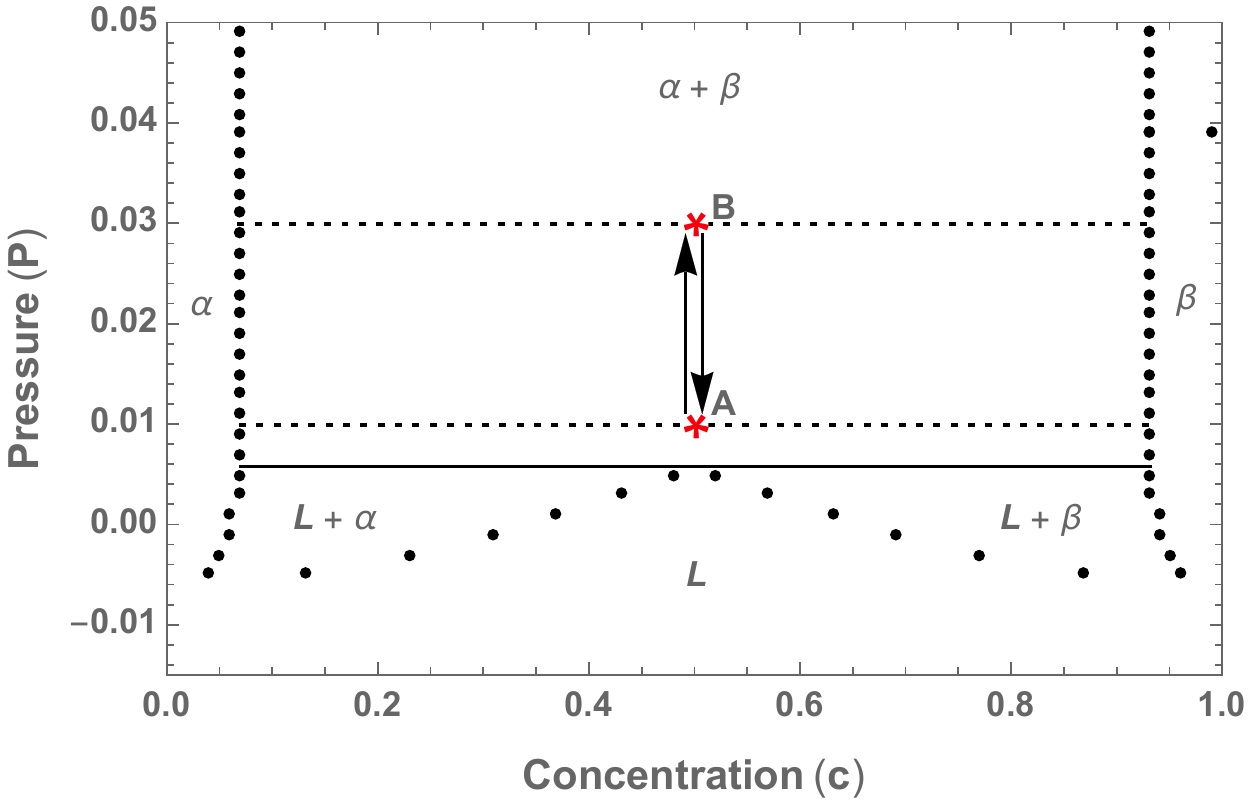}
    \caption{Schematic view of the simulation cycle demonstrating pressure mediated control of pre-melted boundary width, with dashed lines showing the expected equilibrium phases of the system at points A, and B. The system will start with a pressure and total concentration corresponding to point A. After a period of equilibration the system pressure will be quenched to point B. After a further equilibration the system will be quenched back to point A.}
    \label{fig:premeltingPD}
\end{figure}

\begin{figure}[h]
    \centering
    \includegraphics[width=\linewidth]{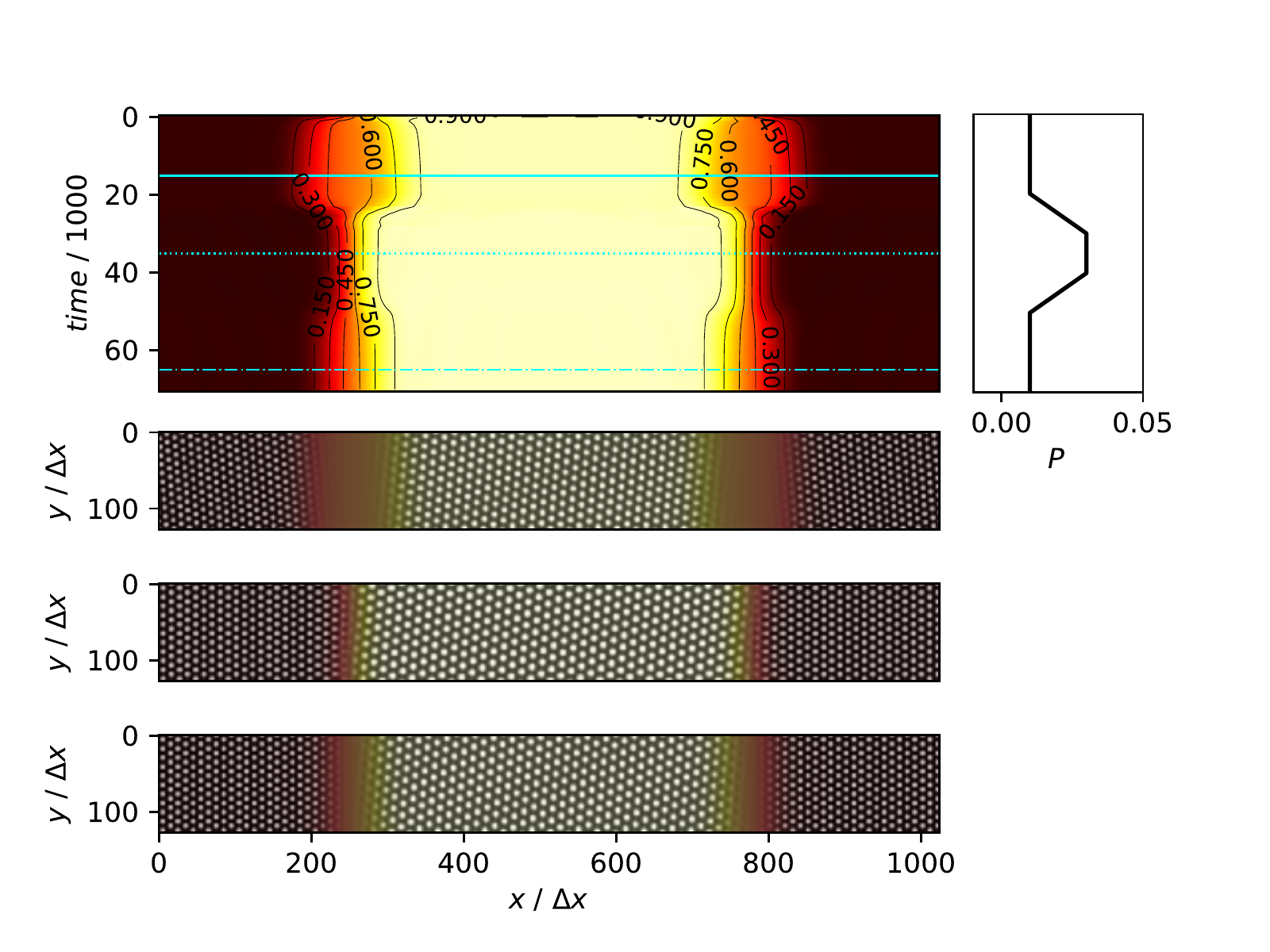}
    \caption{
    (Top left) Concentration vs time of the pressure mediated changes to thermodynamic stability. The concentration shown at each time is the concentration value averaged over the short direction of the channel. (Top right) The grand potential density of the system at the time corresponding to the value in the top left . (Upper middle) The density-concentration profile of the system at the $t=15000$ simulation step represented by the solid horizontal line in the top left figure. At this time  the pressure is at $P=0.01$. (Lower middle) The density-concentration profile of the system at $t=35000$ simulation steps represented by the dotted line in the top left figure. At this time the pressure is at $P=0.03$. (Bottom) The density-concentration profile of the system at $t=65000$ simulation steps represented by the dashed line in the upper left figure. At this time the pressure has once again returned to $P=0.01$.}
    \label{fig:premelting}
\end{figure}

A schematic of the simulation cycle is shown on a constant temperature phase diagram in Fig.~\ref{fig:premeltingPD}. While the results of the simulation are shown in Fig.~\ref{fig:premelting}, where the average concentration profile over time and a selection of density-concentration visualizations are depicted. The target pressure was initially set to $P_0=0.01$ and the system was allowed to equilibrate for 20000 simulation steps. This initial pressure and temperature combination was chosen such that the system was slightly above the eutectic pressure and would therefore exhibit pre-melting.
The target pressure is then quenched to $P_0=0.03$ over 10000 simulation steps and allowed to equilibrate for a further 10000 simulations steps. Under this quench we see that the inter-phase boundary narrows significantly and the amplitude no longer decays to zero between phases. We then quench back to the original pressure of $P_0=0.01$ over a further 10000 simulation steps and allow for a further equilibration of 20000 simulation steps. Having returned to the initial condition, we see that there is once again an increased level of pre-melting, however it is reduced compared to the original amount. This is due to a minor change in the crystallographic orientation of the grains resulting in a change in the misorientation of the grains, which changes the solid-liquid interfacial and the eventual grain boundary energies.
%%%%%%%%

\subsection{\label{subsec:Stability} Control of Thermodynamic Stability}

While most metallurgical processes of casting are performed at atmospheric pressure conditions~\cite[p.~287]{callister}, after casting the treatment of many industrially relevant materials include processes such as annealing and rolling to induce re-crystallization and control grain size and distribution~\cite[Ch.~7]{callister}. With the present model we demonstrate that changes to the pressure that a system experiences can affect the stability of its various phases.

\begin{figure}
    \centering
    \includegraphics[width=0.95\linewidth]{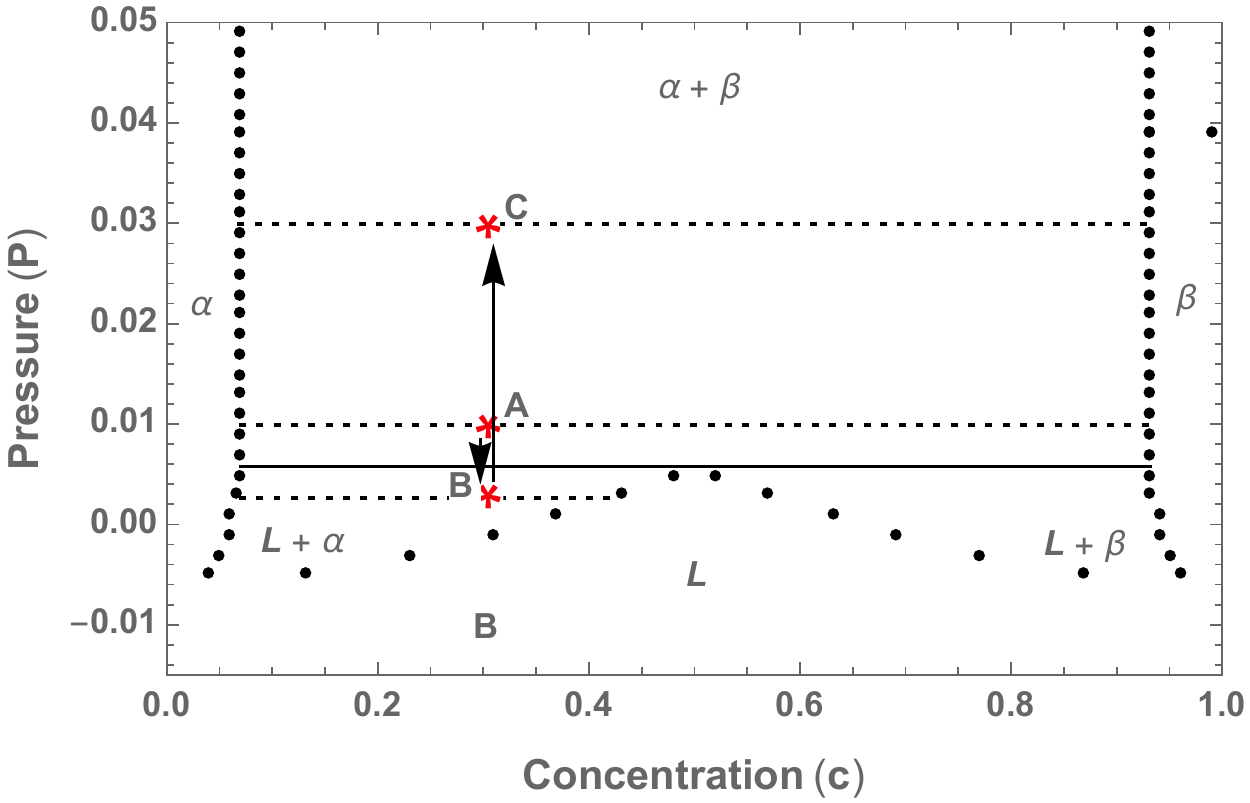}
    \caption{Schematic view of the simulation cycle demonstrating pressure mediated control of thermodynamic stability, with dashed lines showing the expected equilibrium phases of the system at points A, B, and C. The system will start with a pressure and total concentration corresponding to point A. After a period of equilibration the system pressure will be quenched to point B. After a further equilibration in which the beta phase completely melts the system will be quenched to point C; over the course of the quench the beta phase nucleates from the solid-liquid front. After the quench is completed the system is once more allowed to equilibrate for some time.}
    \label{fig:stabilityPD}
\end{figure}

\begin{figure}[h]
    \centering
    \includegraphics[width=\linewidth]{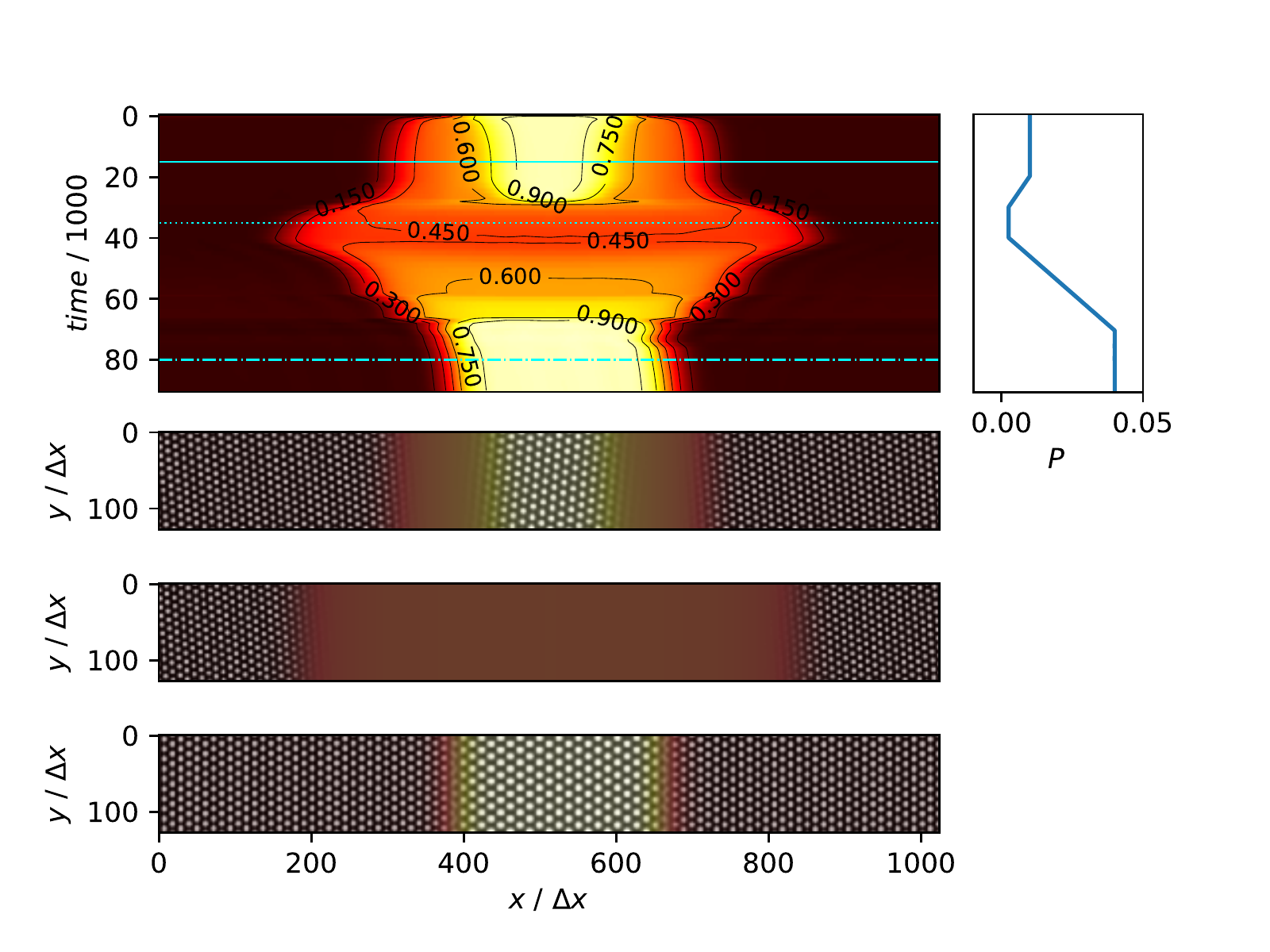}
    \caption{(Top left) Concentration vs time of the pressure mediated changes to thermodynamic stability. The concentration shown at each time is the concentration value averaged over the short direction of the channel. (Top right) The grand potential density of the system at the time corresponding to the value in the top left figure. (Upper middle) The density-concentration profile of the system at $t=15000$ simulation step, as represented by the solid horizontal line in the top left figure. At this time  the pressure is at $P=0.01$. (Lower middle) The density-concentration profile of the system at $t=35000$ simulation steps as represented by the dotted line in the top left figure. At this time the pressure is at $P=0.0025$. (Bottom) The density-concentration profile of the system at $t=80000$ simulation steps as represented by the dashed line in the upper left figure. At this time the pressure is at $P=0.04$.}
    \label{fig:stability}
\end{figure}

To ensure phase elimination we consider a system at $\langle c\rangle = 0.3$, such that there is a preferred fraction of a particular phase. A schematic of the simulation cycle is shown on a constant temperature phase diagram in Fig.~\ref{fig:stabilityPD}. Results are observed in Fig.~\ref{fig:stability}, where we show the average concentration profile over time and a selection of density-concentration visualizations. The target pressure is initially set to $P_0=0.01$ and the system is allowed to equilibrate for 20000 simulation steps. The target pressure is then quenched to $P_0=0.0025$ over 10000 simulation steps before being allowed to equilibrate for a further 10000 simulation steps. Over the course of this pressure quench and equilibration, it is evident that both phases undergo melting, however the $\beta$-phase melts completely due to the quench being below the eutectic pressure.
This process is then reversed by a pressure quench to $P_0=0.04$ over 30000 simulation steps. At some point after the simulation has surpassed the original pressure of $P_0=0.01$ the system precipitates the $\beta$-phase from the over-saturated liquid. This hysteresis is due to the metastability of the over-saturated liquid and the fact that this simulation was performed without noise. The eventual precipitation is mediated by the correlation length of the $\alpha$ phase penetrating into the liquid --- i.e. the solid phase heterogeneously nucleates from the solid-liquid front.
%%%%%%%%%

\subsection{\label{susbec:BoundaryMovement} Non-Equilibrium Inter-Phase Boundary Motion}

In this sub-section we demonstrate strain-induced deviation from the equilibrium phase diagram and the resultant non-equilibrium inter-phase boundary motion that occurs due to this process. It is noted that as with all structural PFC models, this current model does not presently have an explicit coupling between the pressure of the system and the lattice parameter of the crystal. As such, if the crystal is allowed to relax to its proper lattice parameter at one pressure it will not be able to both maintain its proper lattice parameter and simulation box coherency should the pressure change, barring crystallographic rotations or re-crystallization.

In the simulations presented in this sub-section, we consider an overall system concentration of $\langle c\rangle = 0.4$ and excess free energy parameters $k^{(A)}_{10} = 4\pi/\sqrt{3}$, 
$\sigma^{(A)}_{10} = 0.8$, $T^{(A)}_M = 1.0$, $k^{(B)}_{10} =1.1 k^{(A)}_{10}$, $\sigma^{(B)}_{10} = 0.8$, and $T^{(B)}_M =1.0$ such that the increased overlap between peaks will result in larger solubility. 
%%%%%%%
\begin{figure}
    \centering
    \includegraphics[width=0.95\linewidth]{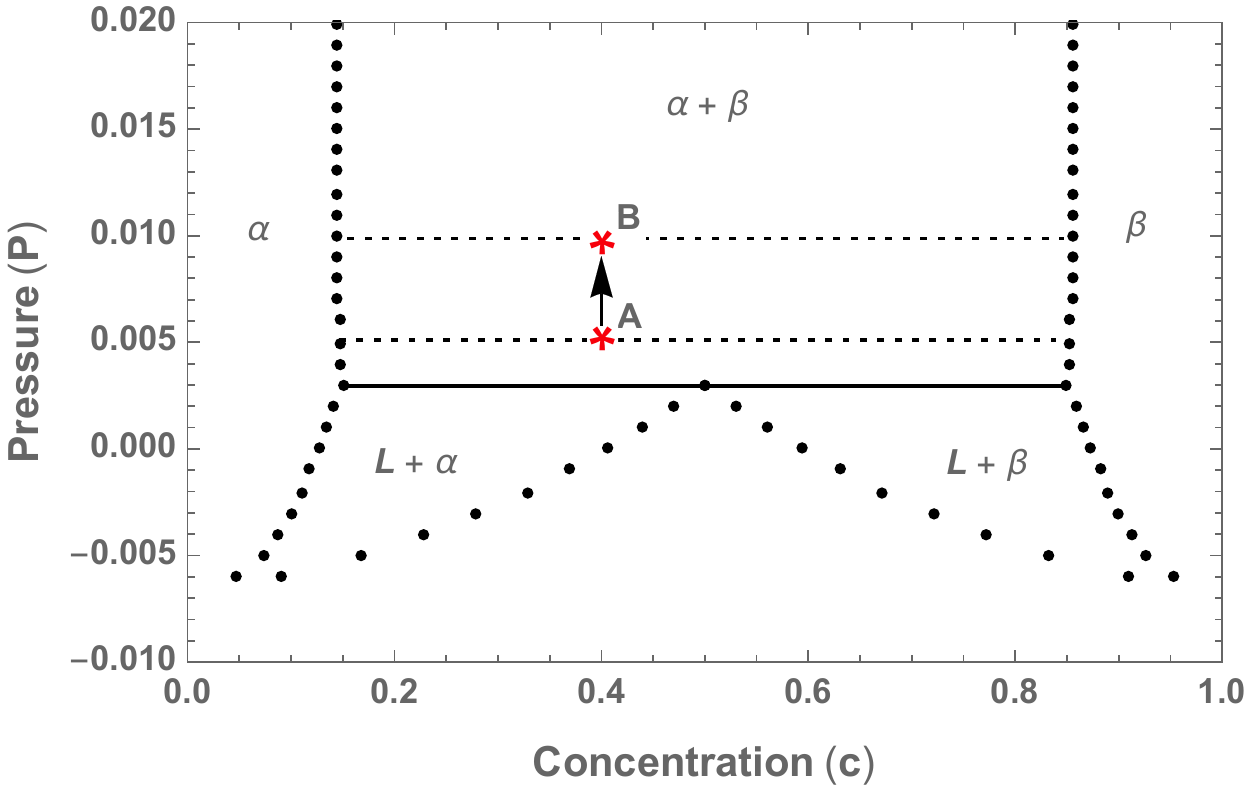}
    \caption{Schematic view of the simulation cycle demonstrating pressure mediated control of non-equilibrium inter-phase boundary movement. The system will start with a pressure and total concentration corresponding to point A. After a period of equilibration the system pressure will be quenched to point B and allowed to equilibrate. Strain on each phase of the bi-crystal will result in non-equilibrium concentrations.}
    \label{fig:GBmovePD}
\end{figure}

\begin{figure}[h]
    \centering
    \includegraphics[width=\linewidth]{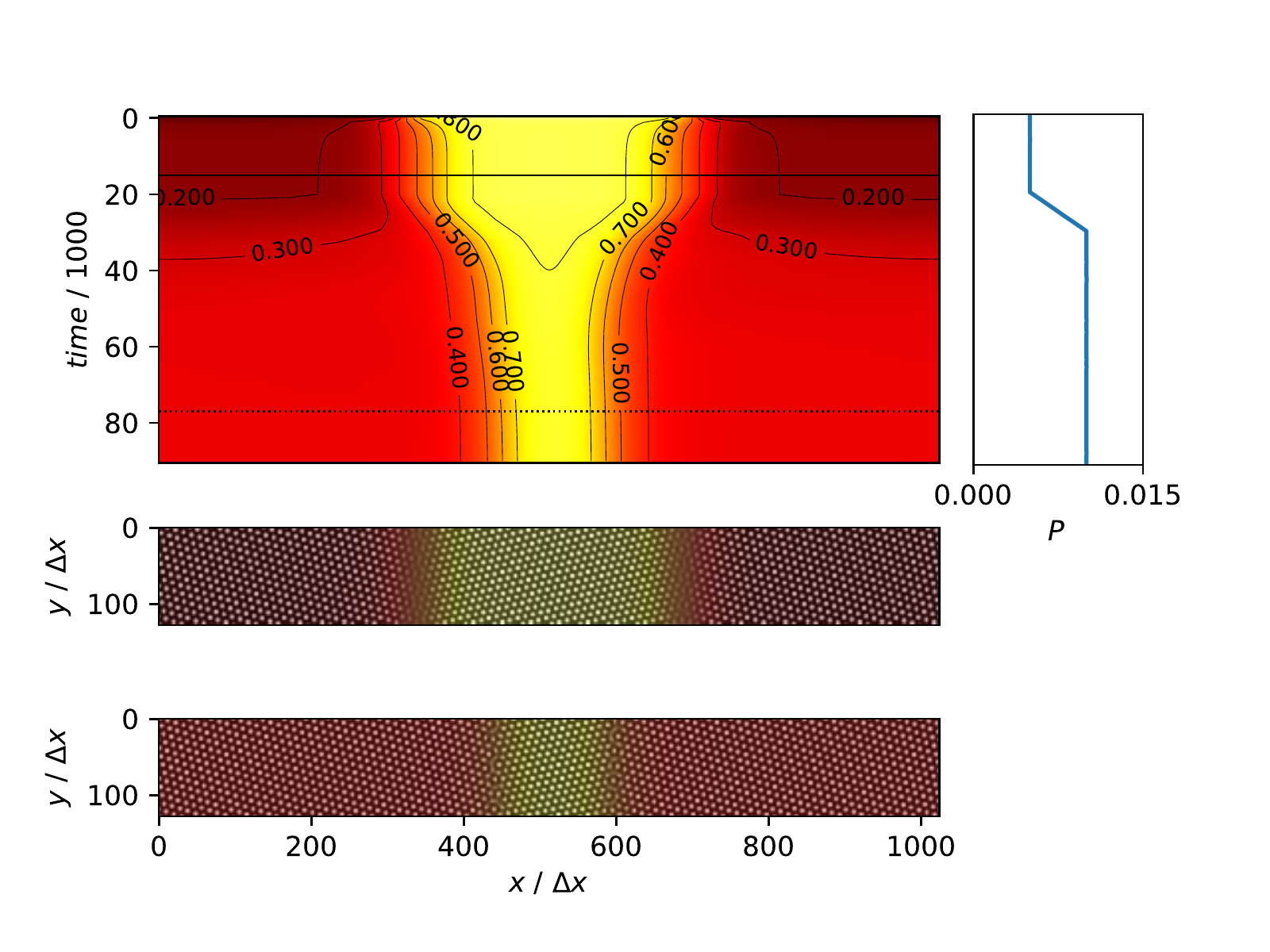}
    \caption{(Top left) Concentration vs time of the pressure mediated inter-phase boundary movement. The concentration shown at each time is the concentration value averaged over the short direction of the channel. (Top right) The grand potential density of the system at the time corresponding to the value in the top left figure. (Middle) The density-concentration profile of the system at $t=15000$ simulation step, as represented by the solid horizontal line in the top left figure. At this time  the pressure is at $P=0.005$. (Bottom) The density-concentration profile of the system at $t=77000$ simulation steps as represented by the dotted line in the top left figure. At this time the pressure is at $P=0.01$.}
    \label{fig:boundaryMovement}
\end{figure}

A schematic of the simulation cycle is shown on a constant temperature phase diagram in Fig.~\ref{fig:GBmovePD}. We show the results of the simulation in Fig.~\ref{fig:boundaryMovement}, where the average concentration profile over time and a selection of density-concentration visualizations are displayed. The target pressure was initially set to $P_0=0.005$ and the system was allowed to equilibrate for 20000 simulation steps. The target pressure was then quenched to $P_0=0.01$ over 10000 simulation time steps before being allowed to equilibrate for a further 60000 simulation time-steps. 
The pressure quench and resultant change in volume took the system entirely off equilibrium due to box strain effects.  As the $\alpha$-phase in this system has the larger lattice parameter, it is energetically favourable for this phase to take on additional solute, lowering its effective lattice parameter and reducing strain. Due to the limited amount of solute in the system the $\beta$-phase has two options --- lower its volume fraction or lower its concentration. As lowering its concentration would increase its lattice parameter and therefore its strain, the energetically favourable choice is to lower its volume fraction and thus the inter-phase boundary moves.

%%%%%%%%%

\section{\label{sec:SoluteDrag} Stress induced grain boundary motion and solute drag}

Control of the growth and coarsening of grains in a solid state material is of particular practical interest due to the intimate relationship between the distribution of grain size and macroscopic material properties~\cite[Ch.~7]{callister}. One of the means by which one can control coarsening is by the addition of various solutes which are attracted to the grain boundaries and have been long known to drag the motion of these boundaries~\cite{cahn1962,hillert1976}. While some work has been done in PFC modelling to study solute drag on moving grain boundaries~\cite{greenwood2012}, this work relied on artificial driving forces through ad-hoc impositions of orientation biases. 

\begin{figure}[h]
    \centering
    \includegraphics[width=0.95\linewidth]{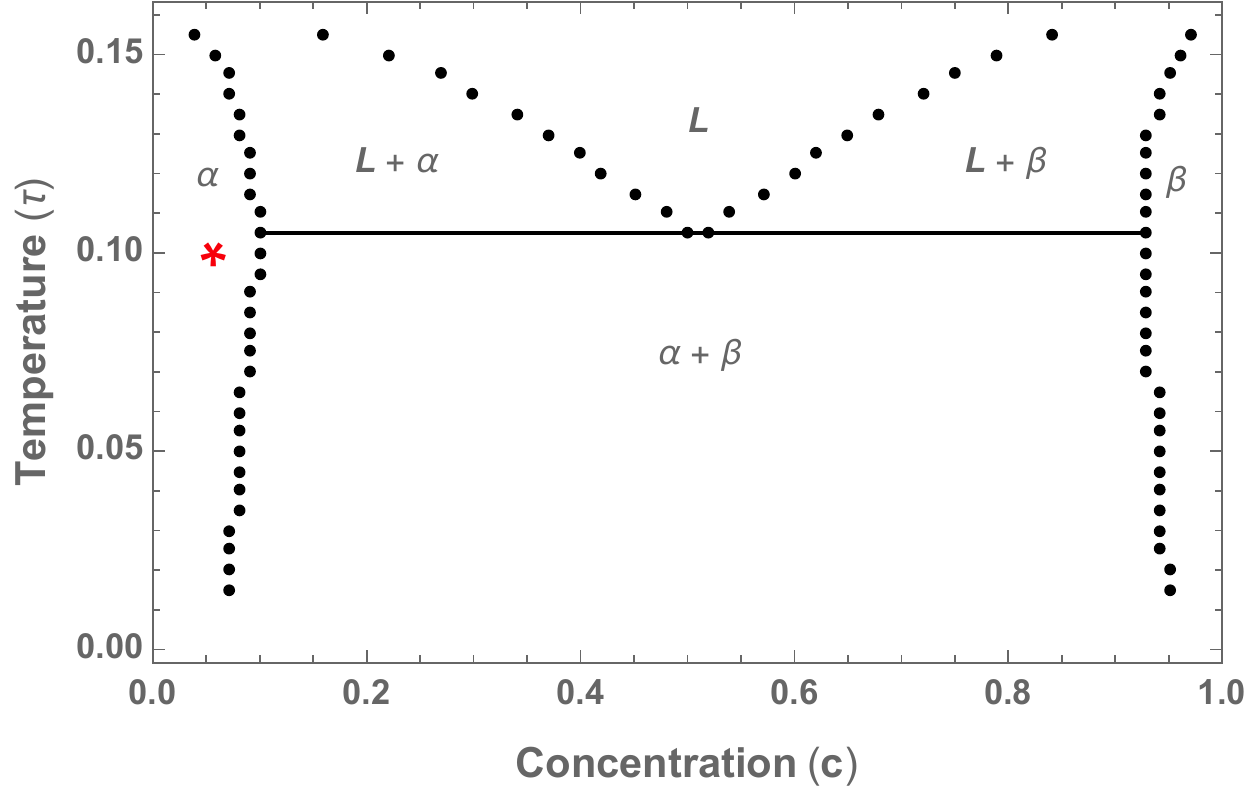}
    \caption{Phase diagram for the triangle-triangle system with a constant model pressure of $P = 0.05$. The parameters used in the ideal free energy were $t=1.4$, $v=1.0$, $w=0.02$, $c_0=0.5$, and $\epsilon(T)=0$ for simplicity. The correlation kernel for the excess free energy used parameters of $k_{10}^{(A)} = 4\pi/\sqrt{3}$, $\sigma_{10}^{(A)}=0.6$, $T_M^{(A)}=1.0$, $k_{10}^{(B)}=2\pi$, $\sigma_{10}^{(B)}=0.8$, and $T_M^{(B)}=1.0$. The $k=0$ peak which controls the bulk modulus of the system was set to a depth of $B_x=1.0$ and width of $\sigma_0=0.8$ for all phases. The red star denotes the $\langle c\rangle$ and $\tau$ values at which the simulation will occur.}
    \label{fig:PhaseDiagramSD}
\end{figure}

In this section we demonstrate as a proof of concept the strain induced stress driven grain boundary motion under isobaric conditions for the purposes of studying solute drag. Similar to the simulations discussed in Sec.~\ref{sec:Kinematics}, we will consider a binary system where both the solute and solvent prefer a triangular lattice symmetry. For completeness, we added a $k=0$ mode to the correlation function here in order to control the bulk modulus of our phases and the density jump between them. A depth, $B_x=1.0$, and width, $\sigma_0=0.8$
of the Gaussian controlling the bulk modulus was chosen to be a constant between phases for simplicity; additionally, the depth of this well was chosen to be temperature independent. The correlation kernel with this addition takes the form
\begin{equation}
    \tilde{C}_{i}(k) = -\frac{B_x}{\tau} e^{-\frac{k^2}{2\sigma_0^2}} + e^{-\frac{T}{T_M}}e^{-\frac{\left(k-k_i\right)^2}{2\sigma_i^2}}\;,
\end{equation}
where for convenience the peak widths have been chosen to have low degrees of overlap.
Some care must be taken in selecting either the width of this Gaussian or the cutoff wavelength of the density-smoothing kernel. Should the cutoff wavelength of the smoothing kernel be significantly larger than the width of the $k=0$ mode of the correlation kernel oscillatory behaviour in concentration can result
on wavelengths that are considered long by the smoothing kernel but not penalized by the correlation function. The two-point correlation function of a real material displays a low-$k$ shelf which smoothly transitions to the first peak of the structure. In the XPFC formalism, correlation kernels can have a plateau between this shelf and the first peak --- it is this control over parameters that results in this oscillatory behaviour for sufficiently poor parameter choices. Thus, reducing the extent of this plateau leads to a reduction of undesirable oscillations in concentration.
The phase diagram for this system is shown in Fig.~\ref{fig:PhaseDiagramSD} with the thermodynamic parameters discussed in the figure caption. 
%%%%%%%%%

To generate a system with a driving force between two crystals of the the same phase, we first generate two super-cells for the crystals using a $64 \times 64$ pixel grid with rectangular volume
element with initial size and constant temperature of $\tau=0.1$, reference temperatures $T_0=T_M=1.0$,
and average concentration of $\langle c\rangle=0.05$ (a generalization of the number conserving feedback loop to a rectangular volume element is discussed in Appendix~\ref{app:Anisotropy}). In this case the super-cells correspond to two unstressed boxes  with a commensurate crystals at orientations of $\theta=0.0^\circ$ and with initial volume elements $\Delta x(t=0) = 0.12487858552217967$ and $\Delta y(t=0) = 0.14419736993450027$ and $\theta\approx3.67^\circ$, respectively, with respect to the short axis of the eventual channel, with initial volume elements $\Delta x(t=0) = 0.12099641271809497$ and $\Delta y(t=0) = 0.13971462290754239$.
Just as they do not start equal, the volume elements of the two super-cells are not equal post relaxation. Thus, seeding the density profile of one of the super-cells into a box with differing dimensions imposes a stress on that crystal. We use this concept to seed the initial condition of the solute drag channel; we set the volume elements of the simulation to that of the relaxed $\theta=0.0^\circ$ super-cell and copy the density profile of the $\theta=0.0^\circ$ super-cell to one part of the simulation domain and the density profile of the $\theta\approx3.67^\circ$ super-cell to the remainder of the domain. 
This set-up corresponds to an approximately isotropic strain on the box-misaligned crystal of $\approx3\%$.

Once this system has been seeded we allow it to reach as close to an equilibrium as it can achieve without changing the physical structure of the system at any given position. To facilitate this "pseudo-equilibration" we run dynamics on the long wavelength components of the free energy using model A dynamics.
\begin{equation}
    \frac{\partial (\chi*\rho_i)}{\partial t} \approx -M_i\left(\chi*\frac{\delta F}{\delta\rho_i}\right)
\end{equation}

\begin{figure}[h]
    \centering
    \includegraphics[width=\linewidth]{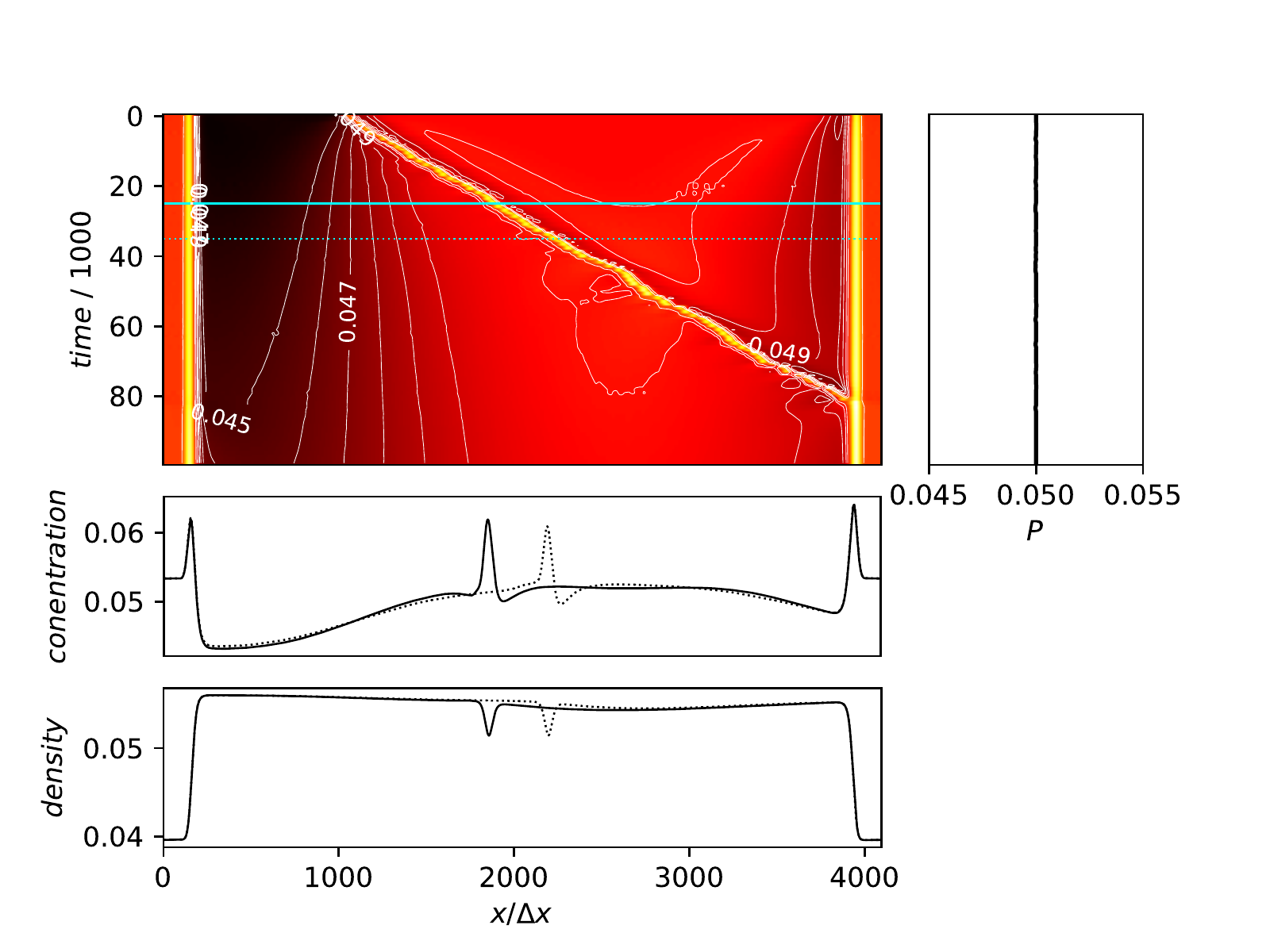}
    \caption{(Top left) Concentration vs time of the strain induced stress driven grain boundary motion. The concentration shown at each time is the concentration value averaged over the short direction of the channel. (Top right) The grand potential density of the system at the time corresponding to the value in the top left figure. (Middle) The concentration profiles averaged along the short dimension at $t=25000$ and $t=35000$ as represented by the solid and dashed lines in the top left figure respectively. (Bottom) The total density profiles averaged along the short dimension at $t=25000$ and $t=35000$ as represented by the solid and dashed lines respectively.}
    \label{fig:SoluteDrag}
\end{figure}

\begin{figure}[h]
    \centering
    \includegraphics[width=\linewidth]{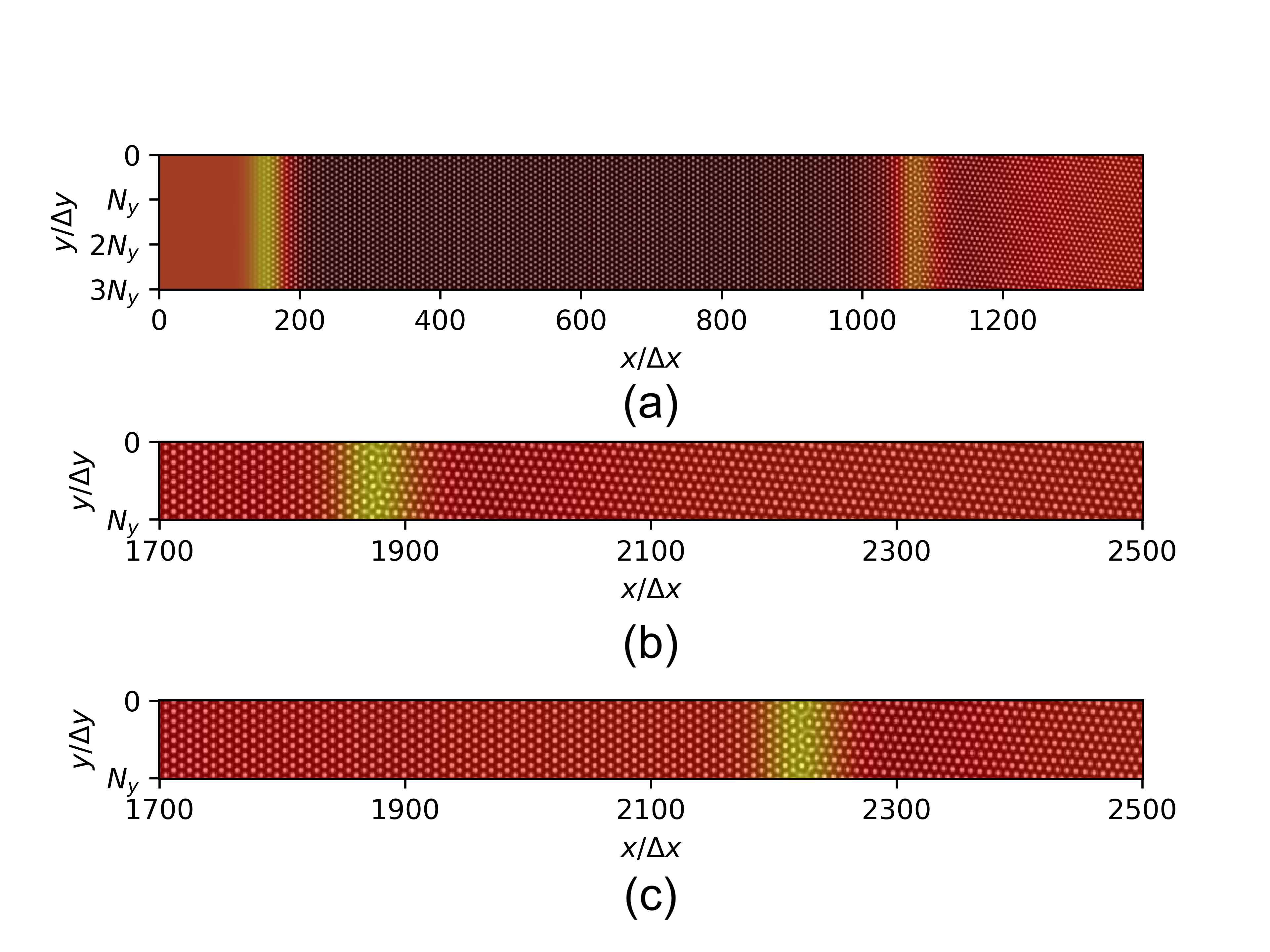}
    \caption{Density-concentration profiles of the system for some selected sub-domain of the system to show various features. (a) The first 1400 pixels in the $x$-direction to demonstrate the liquid penalty layer as well as the system structure at early times, $t=1000$. The $y$-direction is repeated 3 times in this figure for ease of visualization. (b) An 800 pixel snapshot at $t=25000$ as represented by the solid line in Fig.~\ref{fig:SoluteDrag}. (c) The same 800 pixels as (b) but at later time $t=35000$ as represented by the dashed line Fig.~\ref{fig:SoluteDrag}.}
    \label{fig:SDnc}
\end{figure}

Once the pseudo-equilibration is complete we simulate the normal conserved dynamics of each density field at a constant system pressure of $P=0.05$ and density mobilities of $M_A = M_B = 1.0$. To ensure that no crystallographic rotation of the crystal occurs during the simulation we hold the short direction at a constant length, adjusting only the length element in the long direction. Additionally, we reduce the stress in the long direction by utilizing a scheme used by Berry \textit{et al.}~\cite{berry2014} and introduce a penalty function to the edges of the system which generates an artificial liquid. The results of this simulation are shown in Fig.~\ref{fig:SoluteDrag} and Fig.~\ref{fig:SDnc}. The unstressed crystal, being more energetically favourable than the stressed crystal, begins to consume the stressed crystal. Solute is attracted to the grain boundary due to its relatively disordered nature, and thus as the grain boundary moves so too must the solute peak. A thorough examination of the individual component mobilities on the speed of the grain boundary motion will be the topic of an upcoming paper. 
%%%%%%%

\section{\label{sec:Conclusion} Conclusion}

We have presented a relationship that consistently maps the concentration, a long-wavelength field, on to the constituent density fields of a binary alloy, thus making consistent the underlying assumptions used to derive the alloy XPFC model. We have further re-formulated XPFC dynamics in terms of the chemical potentials of the individual component densities, thus allowing for proper conserved dynamics in alloys, as well as independent control of the component mobilities. We have presented a more consistent formulation for the control of pressure than previously done, through volume dynamics and applied these dynamics to an XPFC alloy for the first time, to our knowledge. We have elucidated the means of generating isobaric phase diagrams and dynamical simulations within the XPFC alloy formalism. We tested the fidelity of dynamical simulations against the corresponding phase diagrams. We also demonstrated the ability to control kinetic processes through changes in pressure, an important avenue for controlling phase transformations not previously available in the PFC literature. Finally we have provided a proof of concept for consistently simulating strain induced stress driven grain boundary motion, which could be used for the study of solute drag effects. 
This work also allows various other strain effects to be considered more consistently as shown in Section~\ref{susbec:BoundaryMovement}.

\section{Acknowledgements}

We would like to thank Michael Greenwood for helpful discussions on the topic of solute drag. We also thank The National Science and Engineering Research Council (NSERC) for funding through a strategic grant, and Compute Canada for high performance computing resources. 

\appendix

\section{\label{app:Anisotropy} Rectangular volume elements and anisotropy}

Specializing to two dimensions and generalizing to a rectangular volume element Eqs.~(\ref{eq:NewDensityFlux}) and ~(\ref{eq:NewVolumeChange}) become
\begin{equation}
    \bar{\rho}(x,y,t+\Delta t)=\bar{\rho}(x,y,t)e^{-\frac{M_p(\bar{\omega}-P_0)}{\Delta x\Delta y}},
    \label{eq:RectNewDensityFlux}
\end{equation}
\begin{equation}
    J_\mathrm{isotropic} = -\sqrt{\Delta x(t)\Delta y(t)} + \sqrt{\Delta x(t_0)\Delta y(t_0)\frac{\int_V\mathrm{d}^2r\rho(t)}{\int_v\mathrm{d}^2r\rho(t_0)}}.
    \label{eq:RectNewVolumeChange}
\end{equation}
To further allow for relaxation of any box stresses we also introduce an anisotropic component to the volume flux in the form of an approximated strain energy. Since the ideal contribution should not support any anisotropic stresses
we consider only the excess energy when formulating our phenomenology.
Thus to crudely measure the change in energy of a given deformation of system as a whole we propose the phenomenological but qualitative form
\begin{equation}
\begin{split}
    \epsilon_x &= \frac{1}{V}\int_V\mathrm{d}^2r\,n(r)\mathcal{F}^{-1}\left\{\frac{\partial\tilde{C}_\mathrm{nn}(k)}{\partial k_x}\tilde{n}(k)\right\},\\
    \epsilon_y &= \frac{1}{V}\int_V\mathrm{d}^2r\,n(r)\mathcal{F}^{-1}\left\{\frac{\partial\tilde{C}_\mathrm{nn}(k)}{\partial k_y}\tilde{n}(k)\right\}
\end{split}
\end{equation}
where $\mathcal{F}^{-1}\{...\}$ denotes the inverse Fourier transform. Such a form has a few attractive features for a single bulk phase; it is agnostic to both the structure of the density field and the box dimensions; it vanishes both in the presence of a liquid and an unstressed crystal; most importantly it is easy to calculate numerically. 
As we are mostly interested in the stress present in bulk phases, we have explicitly neglected the contribution of gradients in the concentration, under the assumption that they will be negligible far from grain boundaries, interfaces, and explicit defects. Such an assumption is motivated in part by the fact that the grand potential is only strictly analogous to pressure in the absence of interfaces and defects. In this paper we have found generally good results using the grand potential as a stand-in for pressure across various systems with interfaces; we thus assume that an anisotropy motivated by bulk phases can similarly be applied to coexistence.

To apply this anisotropy the guess-check-correct algorithm is slightly modified. The expected change in volume element is calculated with an anisotropic contribution as
\begin{equation}
\begin{split}
    \delta(\Delta x) &\approx \frac{\Delta t M_p}{\Delta x + \Delta y}(\omega -P_0),\\
    \delta(\Delta y) &\approx \frac{\Delta t M_p}{\Delta x + \Delta y}(\omega -P_0) + \frac{\Delta t M_p \Delta x}{\Delta x + \Delta y}(\epsilon_y-\epsilon_x)
\end{split}
\label{eq:AnisotropicElementAdjust}
\end{equation}
where we have chosen to only apply the anisotropy to the $y$-component so as to not double count the isotropic driving due to the crystal elasticity. In this case the time step is scaled based on the larger of the two expected changes to the volume elements. Next the density is adjusted as per Eq.~(\ref{eq:RectNewDensityFlux}) before changing $\Delta x$ and $\Delta y$ as given by Eq.~(\ref{eq:AnisotropicElementAdjust}). Finally,   Eq.~(\ref{eq:RectNewVolumeChange}) is applied to both $\Delta x$ and $\Delta y$ to ensure proper number conservation.

\bibliography{Bibliography}

\begin{thebibliography}{10}

\bibitem{elder2007}
Ken~R Elder, Nikolas Provatas, Joel Berry, Peter Stefanovic, and Martin Grant.
\newblock Phase-field crystal modeling and classical density functional theory
  of freezing.
\newblock {\em Physical Review B}, 75(6):064107, 2007.

\bibitem{grant2004}
K.~R. Elder and Martin Grant.
\newblock Modeling elastic and plastic deformations in nonequilibrium
  processing using phase field crystals.
\newblock {\em Phys. Rev. E}, 70:051605, Nov 2004.

\bibitem{greenwood2010}
Michael Greenwood, Nikolas Provatas, and J{\"o}rg Rottler.
\newblock Free energy functionals for efficient phase field crystal modeling of
  structural phase transformations.
\newblock {\em Physical review letters}, 105(4):045702, 2010.

\bibitem{greenwood2011}
Michael Greenwood, J{\"o}rg Rottler, and Nikolas Provatas.
\newblock Phase-field-crystal methodology for modeling of structural
  transformations.
\newblock {\em Physical review e}, 83(3):031601, 2011.

\bibitem{greenwood2011binary}
Michael Greenwood, Nana Ofori-Opoku, J{\"o}rg Rottler, and Nikolas Provatas.
\newblock Modeling structural transformations in binary alloys with phase field
  crystals.
\newblock {\em Physical Review B}, 84(6):064104, 2011.

\bibitem{smith2017}
Nathan Smith and Nikolas Provatas.
\newblock Generalization of the binary structural phase field crystal model.
\newblock {\em Phys. Rev. Materials}, 1:053407, Oct 2017.

\bibitem{kocher2015}
Gabriel Kocher and Nikolas Provatas.
\newblock New density functional approach for solid-liquid-vapor transitions in
  pure materials.
\newblock {\em Phys. Rev. Lett.}, 114:155501, Apr 2015.

\bibitem{Wang2017}
Nathan~Smith Nan~Wang and Nikolas Provatas.
\newblock Phase-field-crystal methodology for modeling of structural
  transformations.
\newblock {\em Physical review M}, 1:043405, 2017.

\bibitem{elder2011}
KR~Elder, K~Thornton, and JJ~Hoyt.
\newblock The kirkendall effect in the phase field crystal model.
\newblock {\em Philosophical Magazine}, 91(1):151--164, 2011.

\bibitem{lu2015}
Guang-Ming Lu, Yan-Li Lu, Ting-Ting Hu, and Zheng Chen.
\newblock Phase field crystal study on the phase boundary migration induced by
  the kirkendall effect.
\newblock {\em Computational Materials Science}, 106:170--174, 2015.

\bibitem{greenwood2012}
Michael Greenwood, Chad Sinclair, and Matthias Militzer.
\newblock Phase field crystal model of solute drag.
\newblock {\em Acta materialia}, 60(16):5752--5761, 2012.

\bibitem{fallah2012}
Vahid Fallah, Jonathan Stolle, Nana Ofori-Opoku, Shahrzad Esmaeili, and Nikolas
  Provatas.
\newblock Phase-field crystal modeling of early stage clustering and
  precipitation in metal alloys.
\newblock {\em Physical Review B}, 86(13):134112, 2012.

\bibitem{fallah2013simulation}
Vahid Fallah, Nana Ofori-Opoku, Jonathan Stolle, Nikolas Provatas, and Shahrzad
  Esmaeili.
\newblock Simulation of early-stage clustering in ternary metal alloys using
  the phase-field crystal method.
\newblock {\em Acta materialia}, 61(10):3653--3666, 2013.

\bibitem{fallah2013atomistic}
Vahid Fallah, Andreas Korinek, Nana Ofori-Opoku, Nikolas Provatas, and Shahrzad
  Esmaeili.
\newblock Atomistic investigation of clustering phenomenon in the al--cu
  system: Three-dimensional phase-field crystal simulation and hrtem/hrstem
  characterization.
\newblock {\em Acta materialia}, 61(17):6372--6386, 2013.

\bibitem{elder2010}
KR~Elder and ZF~Huang.
\newblock A phase field crystal study of epitaxial island formation on
  nanomembranes.
\newblock {\em Journal of Physics: Condensed Matter}, 22(36):364103, 2010.

\bibitem{lu2016}
Yanli Lu, Yingying Peng, and Zheng Chen.
\newblock A binary phase field crystal study for liquid phase heteroepitaxial
  growth.
\newblock {\em Superlattices and Microstructures}, 97:132--139, 2016.

\bibitem{seymour2016}
Matthew Seymour and Nikolas Provatas.
\newblock Structural phase field crystal approach for modeling graphene and
  other two-dimensional structures.
\newblock {\em Physical Review B}, 93(3):035447, 2016.

\bibitem{alster2017}
Eli Alster, KR~Elder, Jeffrey~J Hoyt, and Peter~W Voorhees.
\newblock Phase-field-crystal model for ordered crystals.
\newblock {\em Physical Review E}, 95(2):022105, 2017.

\bibitem{cahn1958}
John~W Cahn and John~E Hilliard.
\newblock Free energy of a nonuniform system. i. interfacial free energy.
\newblock {\em The Journal of chemical physics}, 28(2):258--267, 1958.

\bibitem{FFTW}
Matteo Frigo and Steven~G Johnson.
\newblock The design and implementation of fftw3.
\newblock {\em Proceedings of the IEEE}, 93(2):216--231, 2005.

\bibitem{landau1980}
Lev~Davidovich Landau and EM~Lifshitz.
\newblock {\em Statistical Physics, Part 1: Volume 5}.
\newblock Butterworth-Heinemann, 1980.

\bibitem{jugdutt2015}
Bernadine~A Jugdutt, Nana Ofori-Opoku, and Nikolas Provatas.
\newblock Calculating the role of composition in the anisotropy of solid-liquid
  interface energy using phase-field-crystal theory.
\newblock {\em Physical Review E}, 92(4):042405, 2015.

\bibitem{kirkwood1941}
John~G Kirkwood and Elizabeth Monroe.
\newblock Statistical mechanics of fusion.
\newblock {\em The Journal of Chemical Physics}, 9(7):514--526, 1941.

\bibitem{ramakrishnan1979}
TV~Ramakrishnan and M\_ Yussouff.
\newblock First-principles order-parameter theory of freezing.
\newblock {\em Physical Review B}, 19(5):2775, 1979.

\bibitem{Mathematica}
Wolfram~Research{,} Inc.
\newblock Mathematica, {V}ersion 12.0.
\newblock Champaign, IL, 2019.

\bibitem{hsieh1989}
TE~Hsieh and RW~Balluffi.
\newblock Experimental study of grain boundary melting in aluminum.
\newblock {\em Acta Metallurgica}, 37(6):1637--1644, 1989.

\bibitem{adland2013}
Ari Adland, Alain Karma, Robert Spatschek, Dorel Buta, and Mark Asta.
\newblock Phase-field-crystal study of grain boundary premelting and shearing
  in bcc iron.
\newblock {\em Physical Review B}, 87(2):024110, 2013.

\bibitem{straumal2014}
BB~Straumal, A~Korneva, O~Kogtenkova, L~Kurmanaeva, P~Zieba,
  A~Wierzbicka-Miernik, SN~Zhevnenko, and B~Baretzky.
\newblock Grain boundary wetting and premelting in the cu--co alloys.
\newblock {\em Journal of alloys and compounds}, 615:S183--S187, 2014.

\bibitem{callister}
William~D Callister~Jr and David~G Rethwisch.
\newblock {\em Callister's Materials Science and Engineering}.
\newblock John Wiley \& Sons, 2010.

\bibitem{cahn1962}
John~W Cahn.
\newblock The impurity-drag effect in grain boundary motion.
\newblock {\em Acta metallurgica}, 10(9):789--798, 1962.

\bibitem{hillert1976}
Mats Hillert and BO~Sundman.
\newblock A treatment of the solute drag on moving grain boundaries and phase
  interfaces in binary alloys.
\newblock {\em Acta Metallurgica}, 24(8):731--743, 1976.

\bibitem{berry2014}
Joel Berry, Nikolas Provatas, J{\"o}rg Rottler, and Chad~W Sinclair.
\newblock Phase field crystal modeling as a unified atomistic approach to
  defect dynamics.
\newblock {\em Physical Review B}, 89(21):214117, 2014.

\end{thebibliography}
\bibliographystyle{unsrt}

\end{document}